\documentclass[twocolumn, twocolappendix]{aastex631}
\let\tablenum\relax 
\usepackage{amsfonts,amsmath,graphicx,natbib,url,hyperref,nicefrac,siunitx}
\usepackage{amssymb,verbatim,subfigure,paralist,soul,comment, todonotes}
\usepackage[normalem]{ulem} 
\usepackage[dvipsnames]{xcolor}
\hypersetup{colorlinks=true, urlcolor=blue, citecolor=blue}

\newcommand{\Msun}{M_{\odot}}

\newcommand{\ej}{\text{ej}}
\newcommand{\csm}{\text{CSM}}

\newcommand{\FS}{\text{FS}}
\newcommand{\CR}{\text{CR}}
\newcommand{\rRS}{r_{\text{RS}}}
\newcommand{\rFS}{r_{\text{FS}}}
\newcommand{\vFS}{v_{\text{FS}}}
\newcommand{\Eej}{E_{\text{ej}}}
\newcommand{\Mej}{M_{\text{ej}}}
\newcommand{\tcool}{t_{\text{cool}}}
\newcommand{\tsed}{t_{\text{Sedov}}}
\newcommand{\arcsecond}{''}
\DeclareSIUnit\year{yr}
\newcommand{\Rchar}{R_{\text{ch}}}
\newcommand{\Vchar}{V_{\text{ch}}}
\begin{document}

\title{A Novel Parameterization for Rapid Cooling in Supernova Remnants, with applications to the Pa 30 nebula}

\author{Miranda Pikus}
\affiliation{Department of Physics and Astronomy, Purdue University, 525 Northwestern Avenue, West Lafayette, IN 47907, USA}

\author{Paul Duffell}
\affiliation{Department of Physics and Astronomy, Purdue University, 525 Northwestern Avenue, West Lafayette, IN 47907, USA} 

\author{Soham Mandal}
\affiliation{Department of Astronomy, University of Virginia, 530 McCormick Road, Charlottesville, VA 22904, USA} 

\author{Abigail Polin}
\affiliation{Department of Physics and Astronomy, Purdue University, 525 Northwestern Avenue, West Lafayette, IN 47907, USA} 

\begin{abstract}
We systematically study how cooling creates structural changes in supernova remnants as they evolve. Inspired by the peculiar morphology of the Pa 30 nebula, we adopt a framework in which to characterize supernova remnants under different degrees of cooling. Our cooling framework characterizes remnants with a singular parameter called $\beta$ that sets how rapidly the system's thermal energy is radiated or emitted away. A continuum of morphologies is created by the implementation of different cooling timescales. For $\beta \gtrsim 400$, or when the cooling timescale is shorter than $\approx \frac{1}{400}$ of the Sedov time, the ejecta is shaped into a filamentary structure similar to Pa 30. We explain the filament creation by the formation of Rayleigh-Taylor Instability fingers where cooling has prevented the Kelvin-Helmholtz Instability from overturning and mixing out the tips. The ejecta in these filaments have not decelerated and are moving almost completely ballistically at $\approx 95-100\%$ their free expansion speed. In this rapid cooling regime, an explosion energy $\approx \qty{3.5e47}{erg}$ is inferred. We also propose the cooling mechanism required to create these structures necessitates removing energy at a rate of $2\%$ of $E_{\rm ej}/t$, which implies a cooling luminosity of $\approx \qty{e36}{erg\per\s}$. 
\end{abstract}


\section{Introduction} \label{section: Introduction}
The Pa 30 nebula is a very compelling system.  Its recent discovery has been suggested to resolve longstanding tension on linking a known remnant to the historical galactic supernova SN 1181. It hosts a rapidly spinning central object, presumed to be the remains of a white dwarf merger. Sulfur imaging of Pa 30 has traced out radially symmetric filaments surrounding the central star, a configuration never before seen in any known supernova remnant (SNR). With its unusual morphology, Pa 30 serves as means to investigate a new regime of SNR evolution from an observational and theoretical standpoint. 

\subsection{Historical Supernova in 1181} \label{intro subsec: SN 1181}
Ancient Chinese and Japanese records agree on the appearance of a ``guest star" or ``visiting star" in early August of 1181 AD within Cassiopeia, which we will refer to as SN 1181. Based on its observational period, low galactic latitude, and brightness decline rate, SN 1181 records propose it had been a Type I supernova with a peak brightness of around zero magnitudes lasting about six months (\citealt{stephenson_suspected_1971}, \citealt{stephenson_historical_2002}, \citealt{hsi_new_1957}). Historical records identified SN 1181's sky location, associating it with the only known nearby nebula 3C 58 (\citealt{hoffmann_new_2020}, \citealt{stephenson_suspected_1971}, \citealt{hsi_new_1957}). 

However, optical and radio proper motion measurements of 3C 58 demonstrated velocities that were too slow to match up with the transient event SN 1181 (\citealt{van_den_bergh_search_1990}, \citealt{bietenholz_radio_2006}). \citealt{kothes_distance_2013} argued that the link could still be possible but would require a lower supernova brightness, ejecta mass, and surface temperature of the central neutron star. 

\subsection{Discovery of Pa 30 nebula} \label{intro subsec: pa 30 discovery}
In August 2013, the Pa 30 nebula was discovered by Dana Patchick in the Wide-field Infrared Survey Explorer (WISE) archive as part of an amateur astronomer effort to discover and categorize planetary nebulae (\citealt{wright_wide-field_2010}, \citealt{cutri_explanatory_2012}, \citealt{parker_hash_2016}, \citealt{kronberger_new_2016}). Pa 30's location is a degree closer to the assumed location of historical supernova SN 1181 than 3C 58 and its kinematics infer an age agreeable with 1181 AD (\citealt{hoffmann_new_2020}, \citealt{ritter_remnant_2021}). 

WISE imaging in $\qty{22}{\micro\meter}$ characterized the nebula with an infrared circular shell and allowed identification of a central object (IRAS 00500+6713). Optical follow-up revealed a spectrum dominated by broad emission lines of ionized oxygen (\ion{O}{6}) and carbon (\ion{C}{4}). X-ray observations confirmed the ring-like structure and carbon-burning ashes of neon (Ne), magnesium (Mg), silicon (Si), and sulfur (S) enriching both the central object and nebula \citep{oskinova_x-rays_2020}. Spectroscopy also suggests that the central object is very hot with a temperature $\approx \qty{210000}{\K}$ and has extreme outflow speeds $\approx \qty{16000}{\km\per\s}$ (\citealt{gvaramadze_massive_2019}, \citealt{oskinova_x-rays_2020}, \citealt{lykou_new_2023}). 

Carbon-burning ashes and a hot, rapidly spinning central object could be explained by a merging of two white dwarfs (WD). The extreme wind speeds are too high to be generated by radiation alone, but could be accelerated to high enough velocities with the help of a rotating magnetic field (e.g. \citealt{poe_rotating_1989}), which can be created in the wake of a WD merger event (\citealt{beloborodov_magnetically_2014}). \citealt{lykou_new_2023} calculated upper limits for the magnetic field, but those estimates are lower than previously suggested in literature. \citealt{ko_dynamical_2024} also suggested that carbon burning on the surface coupled with a strong magnetic field could create fast winds. There are theoretical WD merger models that are conducive to the progenitor scenario of the central object to some degree. \citealt{loren-aguilar_high-resolution_2009} fits the inferred $\qty{1.5}{\Msun}$ of the central object but under-produces the amount of Si and S. On the other hand, \citealt{kashyap_double-degenerate_2018} demonstrated a ONe-CO WD merger could suffice to produce a central object that enriches its ejecta with Si and S but underestimates the Ne-O fraction \citep{oskinova_x-rays_2020}. 

If a WD merger is the progenitor scenario, a feasible explanation for the SN 1181 transient event is a Type Iax (SNe Iax) (\citealt{oskinova_x-rays_2020}, \citealt{ritter_remnant_2021}, \citealt{yao_observational_2023}). SNe Iax are a group of abnormal Type Ia supernovae (SNe Ia) that are sub-luminous and have slower velocities near peak \citep{foley_type_2013}. Furthermore, \citealt{kashyap_double-degenerate_2018} demonstrated that a ONe-CO WD merger that fails to detonate would expel a small amount of mass, creating a Iax-like event, and leave behind a kicked and spinning super-Chandresekhar WD. Furthermore, a sufficient delay between merger to explosion could diminish overall asymmetries in the resulting nebula \citep{soker_supernovae_2024}.  However, the Pa 30 system still challenges a Type Iax model as explanation for the initial transient. SNe Iax are thought to result in a kicked stellar remnant \citep{kashyap_double-degenerate_2018} but the observed location of Pa 30's central object suggests too slow of an initial kick velocity to make sense \citep{lykou_new_2023}.

\subsection{A Surprising Filamentary Structure} \label{intro subsec: a filamentary structure}
Narrow deep imaging of the ionized sulfur emission [\ion{S}{2}] revealed a peculiar morphology never seen in any known remnant \citep{fesen_discovery_2023}. The Sulfur images mapped out a highly radially symmetric configuration around $\qty{170}{\arcsecond}$ in diameter composed of dozens of long and narrow filaments ranging $\qty{5}{\arcsecond}-\qty{20}{\arcsecond}$ with expansion velocities around $\qty{1100}{\km\per\s}$ (\citealt{ritter_remnant_2021}, \citealt{fesen_discovery_2023}). 

Furthermore, \citealt{cunningham_expansion_2025} presented Integral Field Unit (IFU) observations with the Keck Cosmic Web Imager (KCWI) spectrograph to map the three-dimensional (3D) structure and velocities for an entire radial section of the nebula. This study disclosed key properties of Pa 30:
\begin{itemize}
    \item the time of explosion as $1152^{+77}_{-75} \, \text{yr}$, corresponding to an age of $873^{+75}_{-77}$ yr
    \item a compact distribution of red and blue shifted velocities, ranging from $\qty{600}{\km\per\s}$ to $\qty{1400}{\km\per\s}$
    \item an inner and outer radius bounding the filaments measured to be $r_{\text{in}} \approx \qty{0.6}{pc}$ and $r_{\text{out}} \approx \qty{1.0}{pc}$, interpreted as the forward and reverse shock location
    \item the filaments are traveling almost ballistically with $k = 0.97^{+0.09}_{-0.08}$ where $v = \frac{k r}{t}$ given a velocity $v$, radius $r$, and age $t$
\end{itemize}

\subsection{Interpreting the Filaments} \label{intro subsec: Filaments and RTI}
Understanding SNR evolution requires modeling the interaction of the supernova ejecta with its surrounding material. The freely expanding supernova ejecta collides with the surrounding circumstellar medium (CSM) and a forward and reverse shock are created. As the forward shock propagates, it  sweeps up mass until it begins to decelerate. The energy-conserving behavior that follows is known as the Sedov-Taylor phase. In the shocked region, significant mixing is occuring as fluid instabilities grow at the interface of the contact discontinuity (e.g. \citealt{chevalier_hydrodynamic_1992}). The density gradient between the ejecta and CSM creates the Rayleigh-Taylor Instability (RTI), which sculpts the ejecta into the well-known RTI fingers. In addition, velocity shear at the edges of the RTI fingers gives rise to the Kelvin-Helmholtz Instability (KHI) that typically overturns the heads of the RTI fingers and molds them into mushroom-like shapes. 

The interplay of RTI and KHI creates diverse outcomes in supernova remnants. In the Tycho remnant, RTI creates ejecta resembling cauliflowers (e.g. \citealt{badenes_constraints_2006}). In the Crab nebula, the pulsar-driven wind asymmetrically blows out RTI fingers to create an elaborate web of criss-crossed strands (e.g. \citealt{dubner_morphological_2017}). In Cassiopeia A, the RTI demonstrates its influence in trailing threads of ejecta, but it is important to note since its progenitor was a core-collapse supernova, there are other asymmetries and instabilities at play (e.g \citealt{milisavljevic_jwst_2024}). 

None of these existing examples of RTI in SNRs quite replicate the particular filamentary structure of Pa 30, which has very long, radial filaments that have not been subjected to significant KHI.  However, some planetary nebula do exhibit long radial filaments, which are typically interpreted as  clumps or knots in the ejecta, stretched into long radial structures by strong winds.  This mechanism was suggested by \citealt{fesen_discovery_2023} to explain Pa 30's filaments. Thin lines of material can be seen trailing novae ejecta clumps in GK Per \citep{shara_gk_2012}, windblown tails are formed in novae like DQ Her \citep{vaytet_evidence_2007}, and photoionized filaments can be observed in planetary nebulae like the Helix Nebula (\citealt{odell_cometary_1996}, \citealt{meaburn_nature_1998}). Clumps could replicate the appropriate filamentary structure in Pa 30, but this explanation has caveats. The clumps would need to be distributed uniformly and there is no strong detections of them at either ends of the filaments (\citealt{fesen_discovery_2023}, \citealt{cunningham_expansion_2025}). Furthermore, various studies argue that the forward shock would clear away clumps as it travels outwards (e.g. \citealt{chevalier_hydrodynamic_1992}, \citealt{polin_using_2022}). It has been argued that the strong wind speeds from the central object play a key role in understanding the filaments as well.  In fact, \citealt{coughlin_wind-driven_2025} demonstrated that RTI spikes reminiscent of Pa 30's filaments can be created by modeling the interaction between the central object's shocked wind and surrounding material due to a strong density gradient and low velocity shear.

\citealt{duffell_sculpting_2024} introduced a mechanism that would naturally create the filaments without needing clumps or modeling the wind effects. Two-dimensional hydrodynamic simulations demonstrated efficient cooling diminishes the pressure support in the RTI fingers and limits KHI, preventing the heads from turning over so the fingers elongate into long, thin spikes. The forward shock would be heavily corrugated and the filaments were found to be moving near the free expansion velocity with a ballistic fraction $k \approx 0.9$, in agreement with \citealt{cunningham_expansion_2025}. X-ray modeling of the nebula suggest a similar idea for the inner shocked region, where the predicted cooling and heating plasma timescales were found to be shorter than the dynamical timescales of the shock \citep{ko_dynamical_2024}. This indicates that a large fraction of the shock's internal energy is being radiated away. 

Pa 30 offers an interesting regime of SNR structure to explore the effects that strong cooling of the ejecta has on the resulting kinematics and morphology. Particularly, it gives us an opportunity to build a framework of cooling and RTI finger behavior in mixing of the ejecta in SNRs. Motivated by the results of \citealt{duffell_sculpting_2024}, we systematically explore how cooling can create differences in the remnant structure. Using 3D hydrodynamic calculations, we implement different cooling timescales in a suite of evolving supernova remnant models to investigate the corresponding effects on kinematics, morphology, and observed signatures. In \autoref{section: Methods}, we discuss our numerical methods. In \autoref{section: Results}, we provide the results of our model suite. In \autoref{section: obs markers for pa 30}, we use our results to interpret the Pa 30 nebula and provide observational predictions. Finally, in \autoref{section: Conclusion}, we summarize and discuss the findings of this paper. 

\section{Numerical Methods} \label{section: Methods}

We use the second-order 3D moving-mesh hydrodynamics code \texttt{SPROUT} to evolve our supernova remnant models \citep{mandal_sprout_2023}. \texttt{SPROUT} solves ideal, non-relativistic hydrodynamics on a 3D cartesian grid with a moving mesh methodology proposed by \citealt{springel_e_2010} and \citealt{duffell_tess_2011}. The mesh expands homologously over several magnitudes in time and in size, making it ideal for modeling supernova remnants. It can accurately capture the detailed structure of RTI that forms at the interface of the ejecta and surrounding CSM (\citealt{mandal_measurement_2024}, \citealt{mandal_3d_2023}). 

We use an adiabatic equation of state with an adiabatic index $\gamma = 5/3$ to relate the pressure $P$ and the internal energy density $\epsilon$,
\begin{equation} \label{eq: equation of state}
    P = (\gamma - 1)\epsilon
\end{equation}

\begin{table}[h] 

\centering
\begin{tabular}{||c c c||} 
 \hline
 Symbol & Scaling & Physical Value \\ 
 \hline\hline
 $\Eej$ & -- & \qty{3.5e47}{erg}  \\ 
 $\Mej$ & -- & \qty{0.1}{\Msun} \\
 $\rho_{\csm}$ & -- & \qty{1.5e-25}{\gram \per\cm\cubed}  \\
 \hline \hline
 $\Rchar$ & $\left(\Mej / \rho_{\csm}\right)^{1/3}$ &  $\approx \qty{3.6}{pc}$ \\ 
 $\Vchar$ &   $\sqrt{\Eej / \Mej}$ & $\approx \qty{420}{\km\per\second}$   \\  
 $P_0$ & $\Eej / \Rchar^3$ & $\approx$ \qty{2.6e-10}{erg\per\cm\cubed} \\ 
 $\tsed$ & $\Rchar/\Vchar$ & $\approx \qty{8440}{yr}$  \\ 
 \hline
\end{tabular}
\caption{A list of run-time parameters used in our supernova remnant models. For each quantity listed, we provide a symbol, a scaling relation, and fiducial physical value. This allows us to scale our resulting remnant models in terms of observable properties. In particular, we define a characteristic radius $\Rchar$ and velocity $\Vchar$ that we will be used to quantify the scaling and energetics of our results.
\\
\text{*}In practice, we define code units such that $\Eej = \Mej = \rho_{\csm} = 1$ for the calculation.}
\label{table: runtime params}
\end{table}

We construct a system of code units that can be converted to a set of fiducial values in physical units for comparison to observations. A list of our characteristic system parameters, definitions, scalings, and corresponding physical values is given in \autoref{table: runtime params}. The free parameters that we set and therefore derive the rest of the system properties from are the ejecta energy $\Eej$, ejecta mass $\Mej$, and the circumstellar medium (CSM) density $\rho_{\csm}$. A single time-dependent calculation explores all possible resulting SNR morphologies spanned by these parameters. Thus, \autoref{table: runtime params} displays a combination of $\Eej$, $\Mej$ and $\rho_{\csm}$ that are defined in terms of fiducial values for Pa 30. The fiducial values that we use for scalings are not typical for supernovae. However, the regime of radiative cooling in supernova remnants that we are investigating in this paper is inspired by the Pa 30 system. We use relevant physical scalings for Pa 30 in order to better inform our interpretations (see \autoref{discussion subsec: dynamical match} for details). $\Eej$ is quite sub-luminous for the typical Type Ia population. However, Pa 30 is thought to have risen from a Type Iax, which are fainter white dwarf explosions (e.g. \citealt{foley_type_2013}). We chose a smaller mass for the ejecta of $\Mej = 0.1 \Msun$ based on observational constraints of Pa 30(e.g. \citealt{oskinova_x-rays_2020}). The CSM density $\rho_{\csm}$ is then computed using the ejecta mass and a conservative estimate for the forward shock radius being $\rFS \approx \qty{1.25}{pc}$. In addition to the three free parameters just discussed, we introduce an additional free parameter that controls the efficiency of cooling in the remnant (see \autoref{methods subsec: cooling}). 

In addition, we define the characteristic Sedov time, the amount of time needed for the expanding SNR to sweep up enough circumstellar mass for the reverse shock to cross the ejecta, beginning the transition to the Sedov-Taylor phase of the blastwave's evolution:  
\begin{equation} \label{eq: sedov time}
    \tsed = \frac{(\Mej/\rho_{\csm})^{1/3}}{\sqrt{\Eej/\Mej}} \approx \qty{8440}{yr}
\end{equation}

\subsection{Initial conditions} \label{methods subsec: initial conditions}
We adopt an exponential ejecta profile in density for our supernova remnant models, which has been shown to approximate a typical Type Ia explosion (e.g. \citealt{dwarkadas_interaction_1998}).

For an initial time $t_0$, we model the ejecta density profile as
\begin{equation}
    \rho_{\ej}(r) = \rho_0 \, e^{-r/r_0}
\end{equation}
where 
\begin{equation}
    r_0 = t_0 \sqrt{\frac{\Eej}{6\,\Mej}} 
\end{equation}
and 
\begin{equation}
\rho_0 = \frac{\Mej}{8 \pi \, r_0^3}
\end{equation}

We consider a constant CSM density with small density perturbations distributed uniformly throughout in order to allow the development of fluid instabilities that would occur in nature. We model the circumstellar density in the domain as 
\begin{equation} \label{eq: density perturb}
    \rho_{\csm} (1 + A \sin{k\phi} \sin{k\theta}) \, \, \, \, \, \, \,   \text{where } k = 77, A = .01
\end{equation} 

The values of $A$ and $k$ were chosen to be consistent with low-amplitude perturbations, as in \citealt{mandal_3d_2023}. In that study, the growth of instabilities and structures was found to be independent of the seeded perturbations in the density, for small $A$. This is also demonstrated in \citealt{duffell_rayleightaylor_2017} where the turbulent energy fraction is shown to be independent as well. In \autoref{append: pert seed} we explicitly test the effect of the perturbation strength to demonstrate that it has no effect on the growth of the RTI fingers. In particular, we find the mixing layer width, and therefore formation and structure of the RTI fingers, is not substantially affected by the value of $A$.

We track the distribution of the ejecta and CSM by using a passive scalar where $X = 1$ for pure ejecta and $X=0$ for pure CSM. Once the ejecta density falls below the CSM density, we assume the material at radii after this point as pure CSM. 

We treat the ejecta as freely expanding with a velocity $v = r/t$ and we set both the ejecta and CSM to be very cold with a pressure $P = 10^{-6}\rho V_{ch}^2$. 

\begin{figure*}
    \centering
    \includegraphics[width=\linewidth]{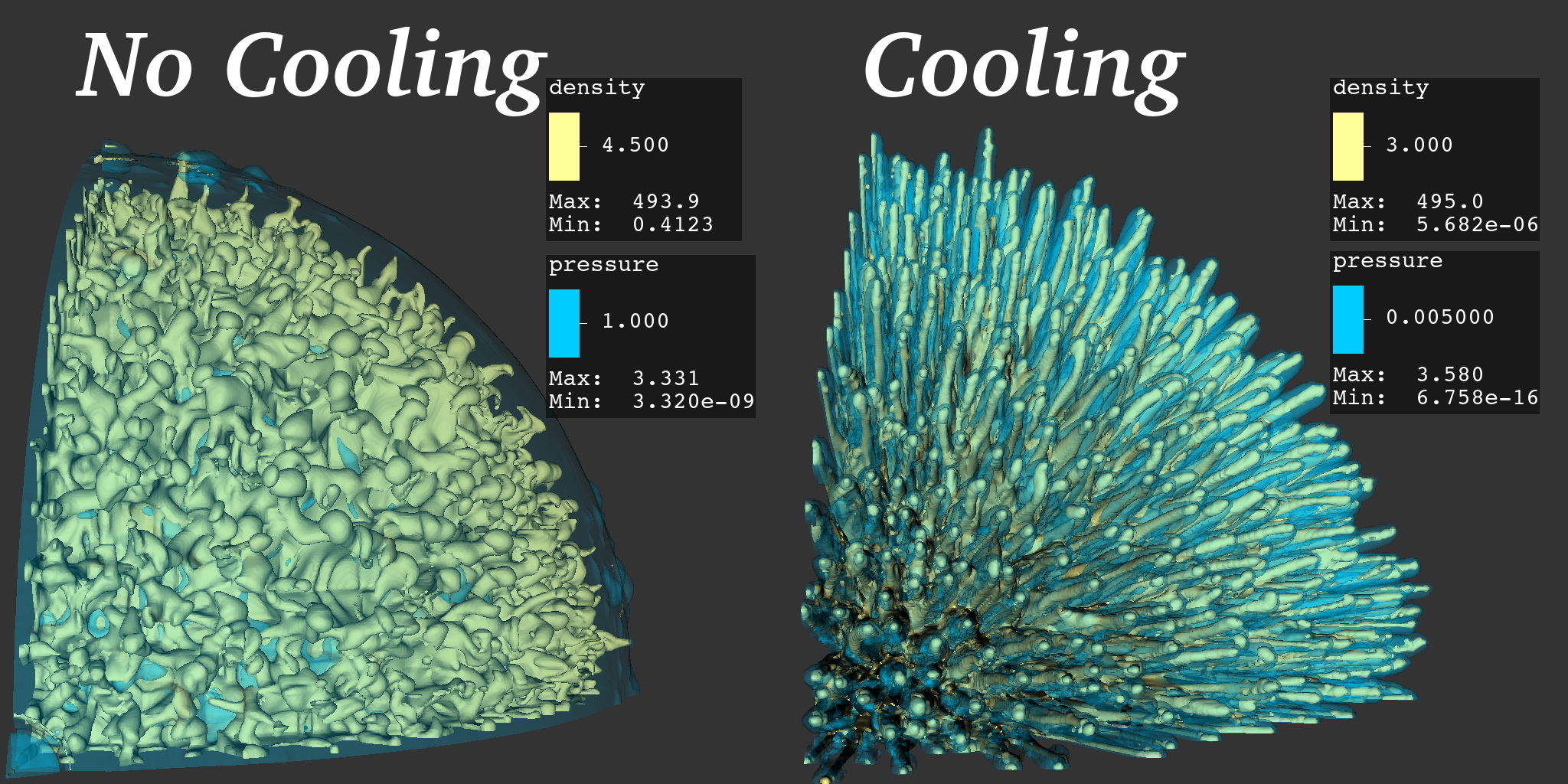}
    \caption{Three-dimensional visualization of the remnant structure for no cooling implemented and for our most rapid cooling with $\beta = 800$. Both models are at $t=0.1 \, \tsed$. The yellow-colored iso-surface is a choice of density intended to target the shape of the ejecta, in units of $\rho_{\csm}$. Furthermore, the transparent blue-colored iso-surface represents the pressure near the forward shock displayed in units of $P_0$. See \autoref{table: runtime params} for physical scalings. With a rapid cooling timescale, the ejecta has been shaped into long and thin spikes that jut out and almost pierce through the forward shock. This causes the contour of the forward shock to be corrugated and sculpted around the filaments.}
    \label{fig:3d isosurface}
\end{figure*} 

\begin{figure*}
    \centering
    \includegraphics[width=\linewidth]{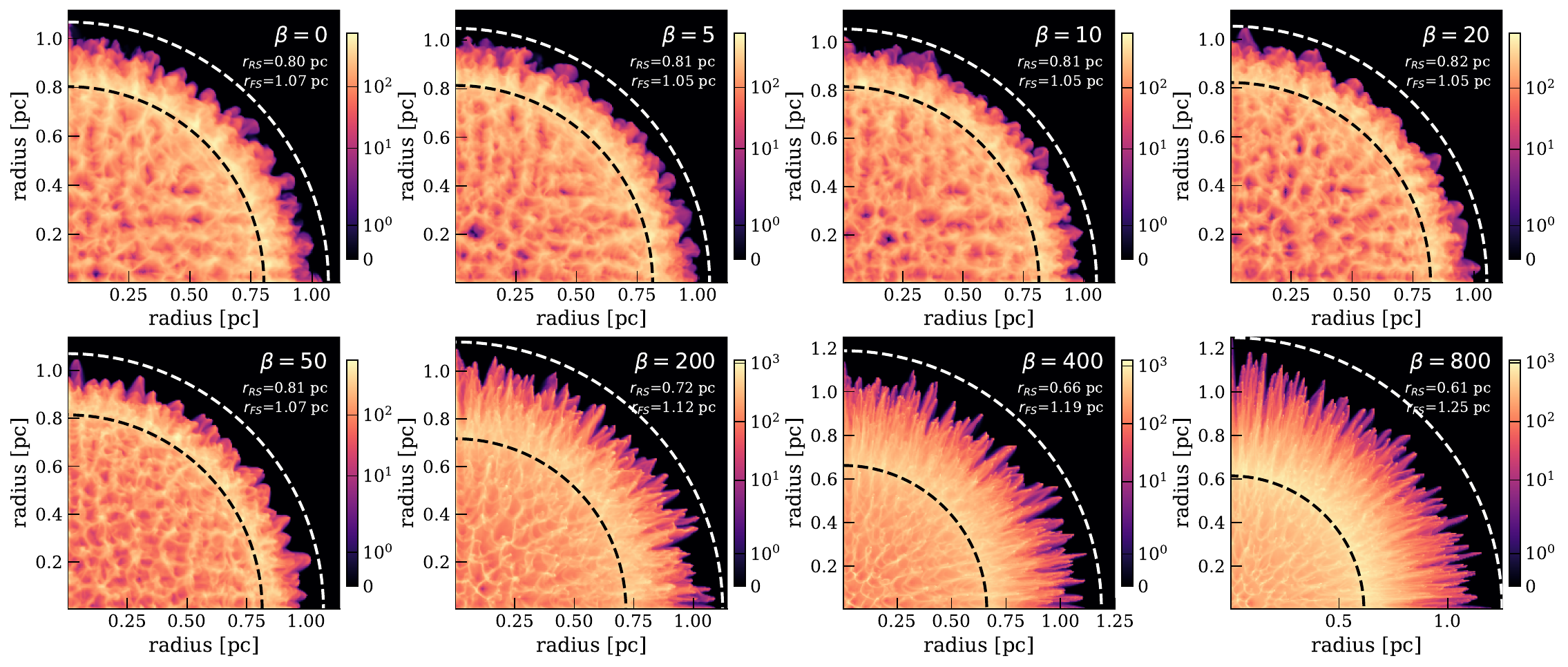}
    \caption{Integrated shocked ejecta density along the $z$ axis for our model suite at $t=0.1 \, \tsed$. The forward shock (white-dashed line) and the reverse shock (black-dashed line) are overlaid for clarity and are determined using the conditions in \autoref{methods subsec: tracking rRS and rFS}. The integrated shocked column density is computed by an integration along the line of sight only in the ejecta at radii greater than the reverse shock radius. The color bar is in units of $\unit{g\per \cm\cubed}$. The projection of the shocked ejecta density demonstrates the extent of radial symmetry the RTI fingers are spatially. For highly radial configurations, the filaments will stack onto each other as opposed to smearing out in a less symmetric case. As $\beta$ increases from the $0$ to $800$, the extent in which the RTI fingers turn over and mix out lessens, resulting in thin and narrow radial filaments.}
    \label{fig: shock ej col density}
\end{figure*}

\subsection{Cooling} \label{methods subsec: cooling}

To mimic cooling in our SNRs, we add a source term to Euler's equation to model a cooling luminosity.

\begin{equation} \label{eq: euler's equations}
    \begin{gathered}
        \partial_t\rho + \nabla \cdot (\rho \boldsymbol{v}) = 0 \\ 
        \partial_t (\boldsymbol{v}) + \nabla \cdot (\rho \boldsymbol{v} \boldsymbol{v} + P \overleftrightarrow{I}) = 0 \\
        \partial_t \left(\frac{1}{2}\rho v^2 + \epsilon \right) + \nabla \cdot \left( \left(\frac{1}{2}\rho v^2 + \epsilon + P \right)\boldsymbol{v} \right) =  S_E\\
    \end{gathered}
\end{equation}

\begin{equation} \label{eq: source term}
    S_E = - \frac{\epsilon}{\tcool} = -\beta \frac{\epsilon}{\tsed}
\end{equation}

We define our source term in terms of the thermal energy $\epsilon$ and a cooling timescale $\tcool$. This quantity can be understood in terms of our characteristic Sedov time $\tsed$ and $\beta$, which we refer to as a dimensionless cooling parameter. $\beta$ is be the prominent free parameter in this study that will allow us to enact any choice of a cooling timescale.

The parameter $\beta$ allows us to systematically explore different thermal cooling rates and thus characterize our parameter space for this study. By assuming cooling is purely related to the thermal energy present in the system, we note that the cooling timescale is proportional to $\frac{\tsed}{\beta}$. For example, $\beta = 100$ mimics cooling at a timescale around $\qty{e-2}{\tsed}$. For $\beta \ll 1$, the calculations approach a usual adiabatic equation of state. For $\beta \gg 1$, we enter a regime of rapid cooling. We interpret $\beta$ as a singular value that determines the constant fraction of thermal energy that is radiated away from the ejecta over time. For the rest of this work, we will discuss the remnant evolution and properties in terms of $\tsed$, for which we scale our SNR dynamics to have an age of $\tsed$ at $t = 1$ in code units.

The nature of this cooling term is agnostic to the type of cooling mechanism, since we do not consider time-dependence or any other quantities besides the thermal internal energy $\epsilon$. Modeling specific cooling mechanisms in supernova remnants would require inclusion of more physics in our source terms. Our primary goal in this work, instead, is to isolate the effects of cooling and understand how this creates structural changes in a supernova remnant as it evolves. We propose our $\beta$-parameterized cooling model as a framework to characterize all SNRs based on how much energy they have radiated away. This parameter space study will aid in narrowing down possible cooling mechanisms and will motivate detailed modeling of specific cooling drivers in the future. 

All of our SNR models are initialized at $t_0=10^{-4}\,\tsed$. The initial domain size is a cube with side lengths of $2 \times 10^{-3}$ in code units, which is approximately $ \qty{0.01}{pc}$ in physical units.  The domain then expands up to a size of about $ \qty{3}{pc}$ during its runtime, following the remnant as it expands. The mesh grid is divided into $256$ zones in each cartesian dimension.

\subsection{Tracking the Forward and Reverse Shock Positions}\label{methods subsec: tracking rRS and rFS}
For each of our models, we also implement an algorithm that tracks the forward and reverse shock radius with time. We locate the reverse shock radius $\rRS$ at the innermost zone such that $v/(r/t) < 0.5$. For the forward shock radius $\rFS$, we locate this at the outermost zone such that the velocity is greater than $5\%$ of the maximum velocity in the entire domain. The accuracy of these conditions will vary throughout time for different $\beta$ values since the evolutionary path for each of the SNRs are very different. However, for most calculations, a sufficient estimate for the reverse and forward shock radius is found with these conditions.

\begin{figure}
    \centering
    \includegraphics[width=\linewidth]{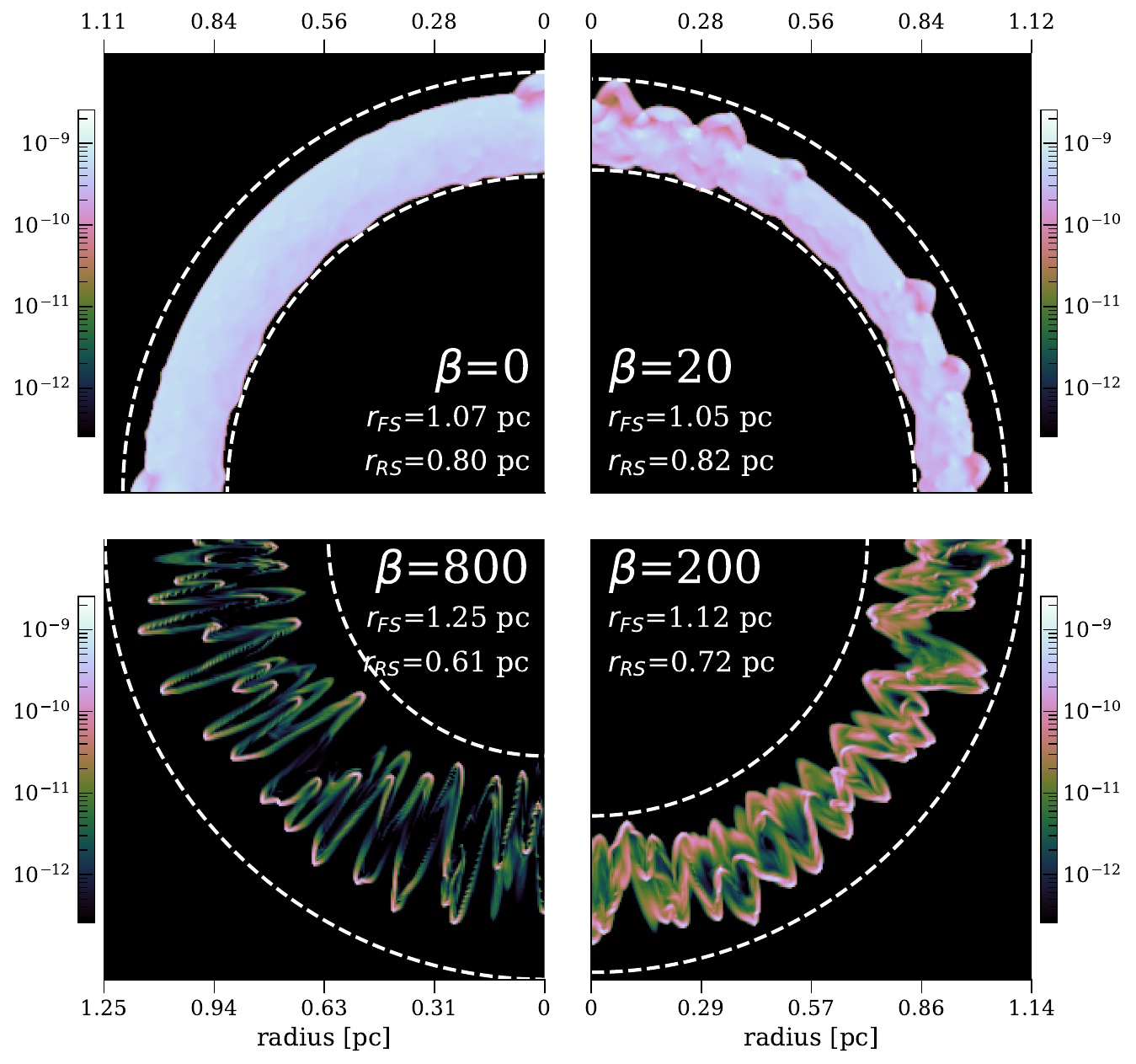}
    \caption{Cross-section slices of pressure ($\unit{erg \per \cubic\cm}$) for four different $\beta$ values. The side lengths are scaled arbitrary to easily compare between different $\beta$ values at $t=0.1 \, \tsed$. The forward shock (outer white-dashed line) and the reverse shock (inner white-dashed line) are overlaid for clarity and are determined using the conditions in \autoref{methods subsec: tracking rRS and rFS}. We suggest that these pressure slices demonstrate the contour of the forward shock, which becomes grooved with increasing $\beta$. In other words, for $\beta=0$, the pressure is distributed uniformly in a smooth spherical shell while the higher $\beta$ runs have a spiked structure that molds around the filamentary structure shown in \autoref{fig:3d isosurface} and \autoref{fig: shock ej col density}. We note that for the more strongly cooled off remnants, the pressure is orders of magnitude less than the standard no-cooling case seen in $\beta = 0$. This could suggest observing SNRs in a regime displayed by $\beta \approx 200-800$ would be difficult as the signals would be much fainter.}
    \label{fig:pressure slices}
\end{figure}

\begin{figure*}
    \centering
    \includegraphics[width=\linewidth]{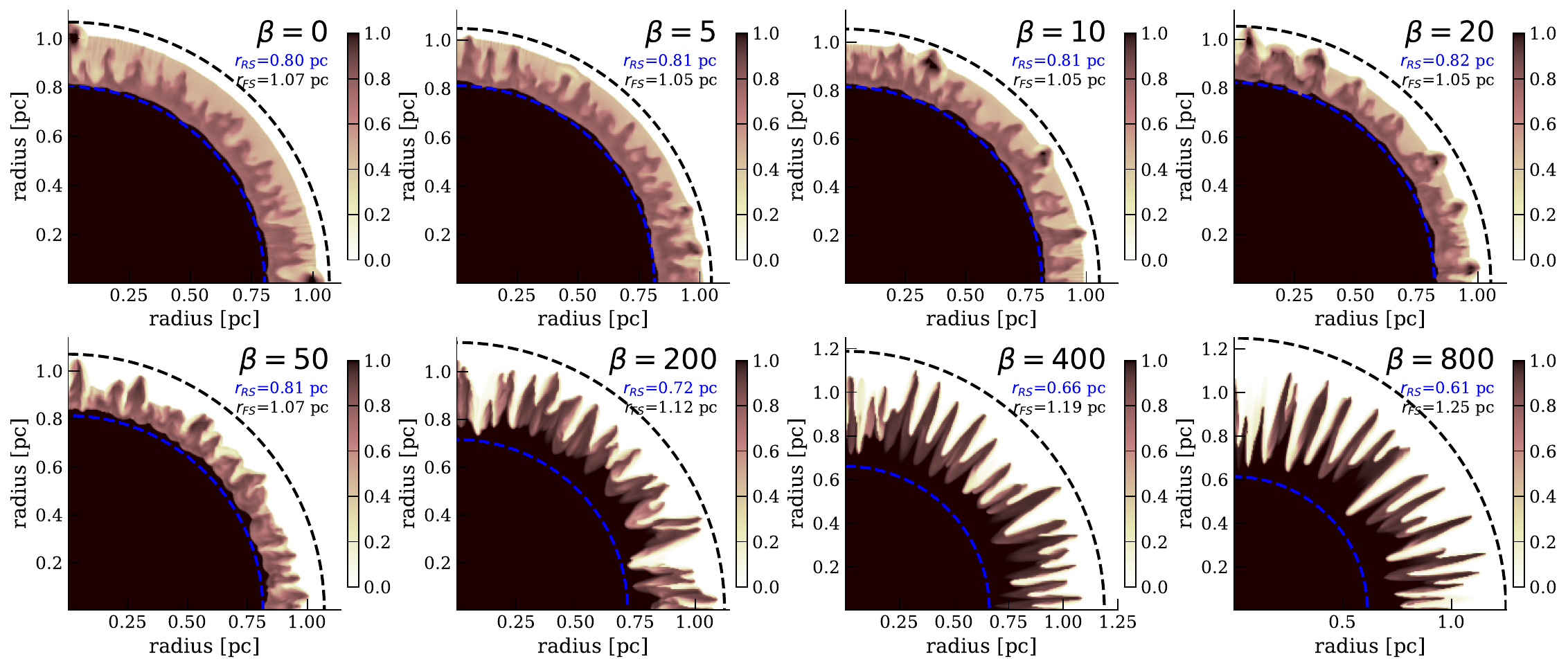}
    \caption{Cross-section slices of the homologous expansion fraction $k = v/(r/t)$ for the velocity in the domain at $t=0.1 \, \tsed$. The distribution of $k$ demonstrates that as $\beta$ increases, or when a more rapid cooling timescale is implemented, the velocity of the material in the filaments gets more ballistic ($k\approx 0.9-1.0$). Stronger cooling prevents the shocked ejecta from decelerating and the Rayleigh-Taylor instability sculpts it into long filaments that have not slowed down from free expansion speeds. The forward and reverse shock positions determined using the conditions specified in \autoref{methods subsec: tracking rRS and rFS} are plotted as dashed blue and black lines respectively.}
    \label{fig: ballistic fraction slices}
\end{figure*}

\begin{figure*}
    \centering
    \includegraphics[width=\linewidth]{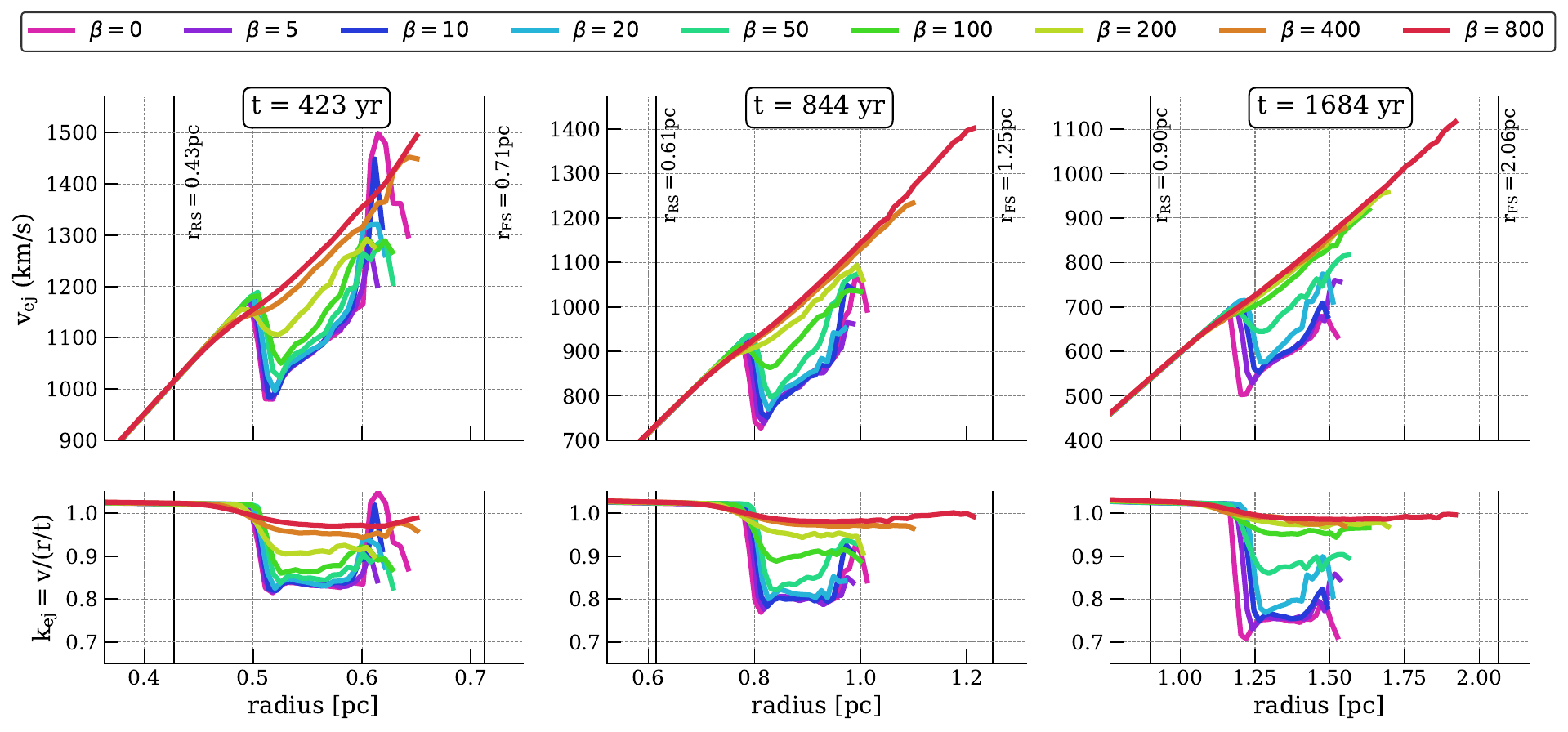}
    \caption{Radially averaged ejecta velocities $v_{\ej}(r)$ and homologous expansion fraction $k_{\ej}(r) = v_{\ej}(r)/t$ for our model suite. Each column denotes a different time. The average is constrained to regions where the ejecta is at least $90 \%$ abundant and weighted by the ejecta density in that region. Each column denotes a different time and the color scheme denotes a different $\beta$. We remark that as $\beta$ is increased, the ejecta velocity becomes more ballistic within the shocked region, with $k \approx 1$ in the most extreme cases. We remark on the continuum achieved in the velocity profiles when comparing between $\beta$s. The forward and reverse shock are denoted as black vertical lines for clarity (see \autoref{methods subsec: tracking rRS and rFS}). We note that the locations of the shocks identified in the radially averaged velocity profiles themselves are different, but we explain this by our averaging process that was weighted by ejecta-dominant zones.}
    \label{fig:vej rad profs}
\end{figure*}

\begin{figure}
    \centering
    \includegraphics[width=\linewidth]
    {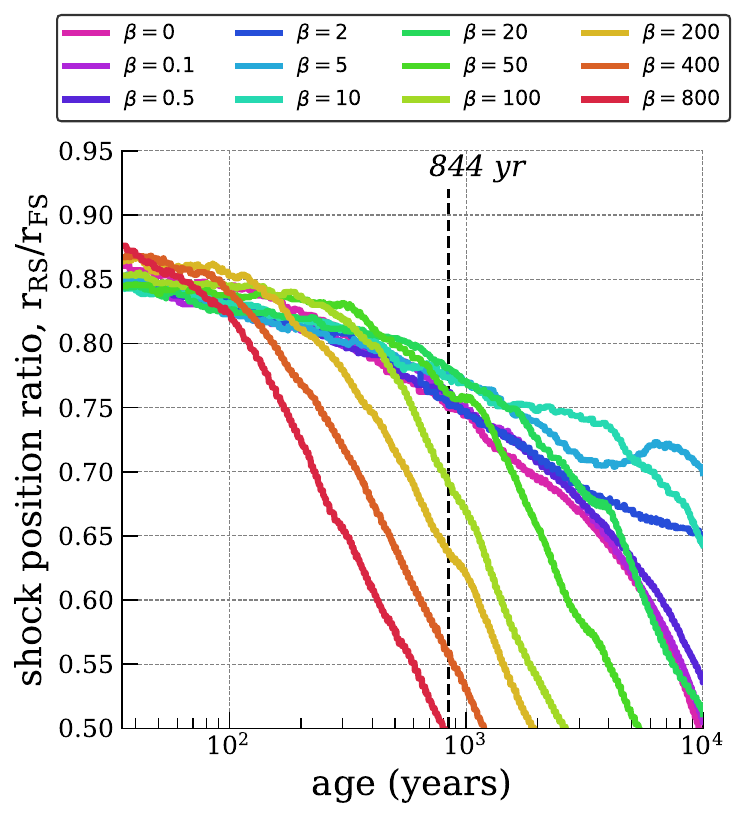}
    \caption{The reverse to forward shock position ratio $\rRS/\rFS$ with time. The color scheme differentiates between the $\beta$ runs. The reverse to forward shock position ratio provides a way to dynamically match our model suite with other systems. As expected, SNRs that are cooled off more efficiently evolve faster (i.e. higher $\beta$ runs show their shock position ratio reaching lower values before the others) since they lose energy more rapidly. We do this for Pa 30 in \autoref{section: obs markers for pa 30}. We mark the assumed age of Pa 30 as $\qty{844}{yr}$ as a dashed black line. We also note that infrared imaging of the nebula estimates a shock ratio of $\approx 0.6$ (see \citealt{cunningham_expansion_2025}).}
\end{figure}

\begin{figure*}
    \centering
    \includegraphics[width=\linewidth]{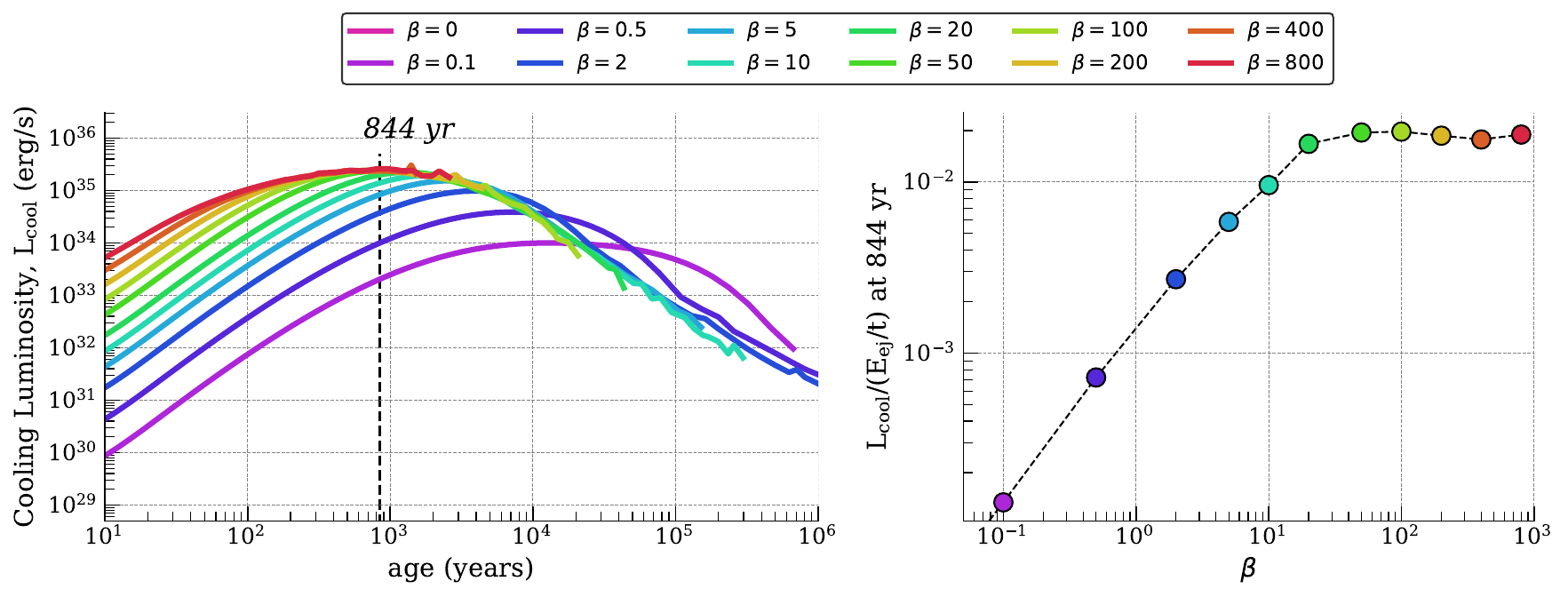}
    \caption{In the left panel, we plot cooling luminosities as a function of time for each $\beta$ model. The cooling luminosity at each time step is computed by summing the total thermal energy in the system multiplied by a factor of $\beta/\tsed$. The cooling luminosity curves rise and fall with a peak value that occurs earlier in time with larger $\beta$s. In the right panel, we plot the ratio of the cooling luminosity $L_{\text{cool}}$ to $\Eej/t$
    at $\qty{0.1}{\tsed} \approx \qty{844}{yr}$ as a function of $\beta$. This plot demonstrates that as our model suite enters a regime of rapid and strong cooling ($\beta \gtrsim 20$), this cooling ratio asymptotes to a maximum value of $ \approx .02$.  Since this is measured at the peak of the luminosity curve, this means that no more than $\approx 2\%$ of the initial ejecta energy could be radiated away by this time.}
    \label{fig:cooling luminosity}
\end{figure*}

\section{Results} \label{section: Results}
We present our hydrodynamical results from our model suite of $\beta$s. We find that a general continuum and convergence of SNR structure is achieved as radiative cooling is increased. The numerical calculations suggest $\beta \gtrsim 400$ serves as an empirical threshold for the Kelvin Helmholtz Instability to be suppressed enough that the Rayleigh-Taylor Instability fingers are molded into long, narrow filaments. $\beta$ is used to explore the continuum and therefore, the value of $\beta$ should be interpreted as a relative regime in which to interpret observed systems, not an exact value. Within this framework, we specifically discuss the morphology, velocity structure, and constraints on corresponding cooling luminosities. 

\subsection{Morphology} \label{results subsec: morphology}
Our model suite of $\beta$ values produces a continuum of morphologies and evolutionary pathways for a SNR. As $\beta$ increases, the SNRs are more rapidly cooled off and the filamentary structure becomes more apparent. \autoref{fig:3d isosurface} displays a 3D rendering of the density and pressure contours for the no cooling case with $\beta=0$ and the most rapid cooling timescale using $\beta = 800$ at $t = 0.1 \, \tsed$. The RTI fingers that are created at the contact discontinuity look starkly different when cooling is implemented. Thermal cooling is efficient enough to suppress the KHI and allow the RTI fingers to morph into long and thin spikes arranged uniformly around the entire sphere. As expected, the pressure is magnitudes smaller for $\beta = 800$ than for $\beta = 0$ since this remnant has lost large amounts of thermal energy. The pressure surfaces overlaid in the 3D renderings demonstrates how the contour of the forward shock is transformed under rapid cooling. For the no-cooling case, the iso-surface of pressure is smooth and spherical and the ejecta has not reached the forward shock. Contrarily, for $\beta=800$, the ejecta filaments have reached the forward shock and sculpted it into a highly corrugated, grooved contour. We observe that our ``no cooling" case appears similar to the morphology of the Tycho SNR, while our most aggressive cooling case with $\beta = 800$ resembles Pa 30. 

This progression of the filament creation and the forward shock corrugation as the SNRs are more rapidly cooled off is exemplified in \autoref{fig: shock ej col density} and \autoref{fig:pressure slices}. In \autoref{fig: shock ej col density}, we show the integrated ejecta column density along the $z$-axis in the shocked region. The shocked region is defined as radii between the forward and reverse shock, which are calculated by the conditions discussed in \autoref{methods subsec: tracking rRS and rFS}. Showing the column density as opposed to a singular cross-sectional slice emphasizes the radial arrangement of the filaments. As $\beta$ increases, the filaments stack onto each other and make the resulting shocked area brighter and concentrated along the radial directions where the filaments lie. 

The distribution of pressure as a function of cooling is portrayed in \autoref{fig:pressure slices} where we show four cross-sectional slices in pressure for $\beta = 0, 20, 200, 800$. As the forward shock travels through the surrounding material, it shocks the CSM and heats it up. Therefore, the pressure slices describe the shape and structure of the shocked region under different amounts of cooling. For the no cooling case ($\beta = 0$), the shocked region is a smooth, spherical shell where pressure is evenly distributed throughout. As $\beta$ increases, the shocked region is morphed into a bumpy structure where the pressure is more concentrated along the outer edge of the shocked region.

The distribution and value of the pressure in the different $\beta$ cases also demonstrate the impact cooling has on the behavior and state of the forward and reverse shocks. For typical SNRs, we expect the pressure to look similarly to the upper left panel in \autoref{fig:pressure slices} for $\beta = 0$, meaning there is a clear demarcation between high pressure and low pressure fluid, suggesting identification of the shock positions are relatively straight forward. However, As $\beta$ is increased, the pressure drops magnitudes lower than the no cooling case and becomes highly corrugated. This result suggests that for remnants undergoing significant cooling in this regime of $\beta$s, the forward and reverse shock would be difficult to directly image.

\subsection{Velocity Structure} \label{results subsec: velocity structure}

The RT fingers form at the contact discontinuity, mixing the shocked ejecta and CSM. Under rapid cooling, the Kelvin-Helmholtz instability is suppressed and the overturning of the tips of the RT fingers is prevented. Thus, the ejecta in the RT fingers morph into long uniform spikes and limited mixing of CSM and ejecta occurs.  CSM is redirected around ejecta filaments, which do not sweep up a substantial amount of mass and do not significantly decelerate.

In \autoref{fig: ballistic fraction slices}, we plot cross-sectional slices for our model suite of the ballistic fraction $k = v/(r/t)$. A value of $k = 1.0$ corresponds to homologous expansion. In the cross-section images, we show how much the ejecta strays from homologous speeds within the shocked regions for different choices of $\beta$. First, the distribution of the ballistic fraction provides another way to portray the emergence of the filaments with higher $\beta$ values. By comparing to \autoref{fig: shock ej col density}, we find that the placement of the filaments is associated with near ballistic velocities and ejecta placement. Second, we observe how the distribution of the ballistic fraction becomes more uniform throughout the filaments, signifying mixing is limited as cooling is increased. In particular, we measure that for $\beta \gtrsim 200$, most of the material inside the filaments is moving at $\gtrsim 90 \%$ their free expansion speed. 

We provide a more detailed examination of the velocity composition in each SNR of our model suite in \autoref{fig:vej rad profs} where we display radially averaged velocity profiles in the ejecta (top panels) and ballistic fraction (bottom panels) around $t = 0.1 \tsed$.  The averaging is weighted by the ejecta density and only accounts for regions in the domain where the ejecta abundance is greater than $90\%$. As $\beta$ increases, the extent of the shocked region bounded between the reverse and forward shock lengthens, which can be explained by the elongation of the RT fingers. Furthermore, the transition of the velocity around the reverse shock becomes more gradual since the material is moving closer to the initial free expansion speed. Generally, a higher $\beta$ results in faster velocities in the shocked region as well. The bottom panel shows that cooling causes the ejecta to move closer to homologous speeds and demonstrates the progression of the material's expansion factor from $\approx 0.8$ to $\approx 1.0$.  

\subsection{Cooling Luminosity} \label{results subsec: Lcool}
The nature of the $\beta$ parameter behaves like a cooling mechanism associated with a timescale $t_{\rm cool} = \frac{1}{\beta} \, \tsed$, which will then have an associated luminosity that might be observable. The energy that is removed via this methodology is purely associated with the thermal energy in the system.  We plot the integrated cooling sink term (\autoref{eq: source term}) at each step in time for each $\beta$ run in the left panel of \autoref{fig:cooling luminosity}. We denote this quantity as $L_{\text{cool}}(t) = \int \epsilon_{\text{th}}(t) \beta /\tsed $. Thus, this can be interpreted as the cooling luminosity corresponding to the cooling mechanism or mechanisms that can be modeled with a timescale $\approx \frac{1}{\beta} \, \tsed$.

In the right panel of \autoref{fig:cooling luminosity}, we plot the estimated fraction of the ejecta energy the cooling luminosity is at the relative age of Pa 30 being $\sim \qty{844}{yr}$ as a function of $\beta$. The right panel demonstrates the percentage of ejecta energy that is converted into cooling luminosity as a function of $\beta$ around the age of Pa 30, $\sim \qty{844}{yr}$. Faster $\beta$ gives higher cooling luminosity $L_{\rm cool}$, until the limit where the cooling timescale is so short that shocked thermal energy is immediately released as it is generated.  In this limit, we measure $L_{\rm cool} \lesssim 0.02 E_{\rm ej}/t$.  Furthermore, by comparing to the morphological structures represented in \autoref{fig: shock ej col density} and \autoref{fig: ballistic fraction slices}, the long filamentary structures are created for $\beta \gtrapprox 400$ and for all values of $\beta$ tested, $L_{cool}/{(\Eej/t)}$ is never more than $2\%$.  In \autoref{discussion subsec: dynamical match}, we will convert this to physical units to give a prediction for the luminosity in erg/s.


\begin{figure}
    \centering
    \includegraphics[width=\linewidth]{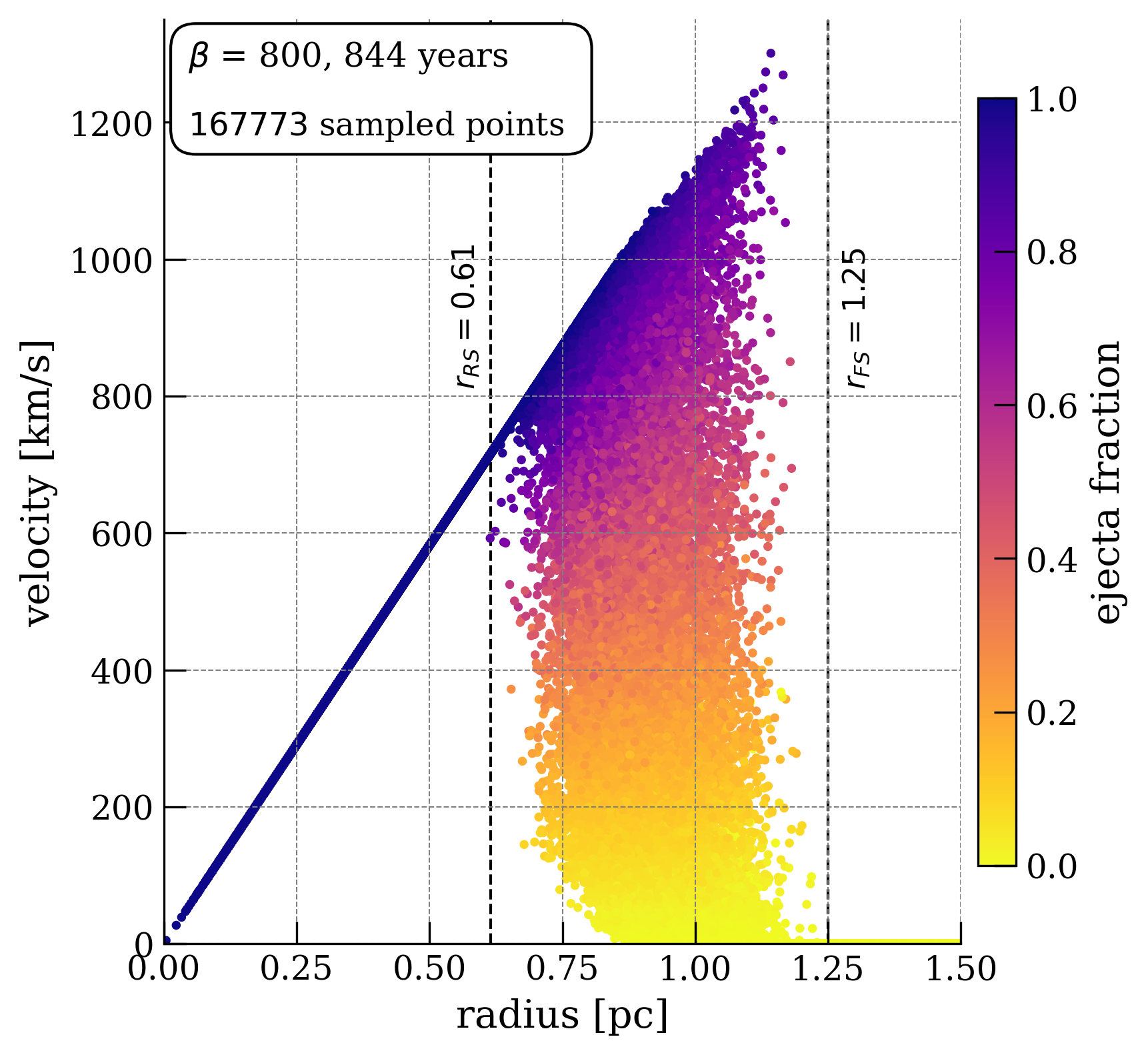}
    \caption{Scatter plot of selected zones for $\beta = 800$ at $t = 0.1 \, \tsed \approx 844$ years. The x-axis is the radius of the selected point and the y-axis is the total velocity for that point. The points are colored by their ballistic fraction the ejecta fraction, or passive scalar value. The forward and reverse shock positions are highlighted by a light grey line, which are found by our estimated tracking algorithm discussed in \autoref{methods subsec: tracking rRS and rFS}. We note that all the zones before the reverse shock are traveling at free expansion speed. At radii greater than the reverse shock, a greater spread of velocities is apparent throughout the entire domain at each radius. As expected, at radii greater than the forward shock, all velocities drop to zero.}
    \label{fig:scatter plot}
\end{figure}

\section{Observational Markers for Pa 30} \label{section: obs markers for pa 30}

Using the results of this work, we provide predictions and suggestions for what could be shaping the filamentary structure of Pa 30.  In particular, we find a best-matched model for Pa 30 within the parameter space spanned by $\beta$, $\Eej$, $\rho_{\csm}$, and $\Mej$.  We also discuss various observational signatures that one could look for in Pa 30 to prove or disprove whether RTI with strong cooling, strong winds, random clumps, or some other mechanism could be responsible for the nebula's morphology.

\subsection{Dynamical Match for Pa 30} \label{discussion subsec: dynamical match}
In this section, we choose the best dynamical match in our model suite to Pa 30 by invoking morphological and velocity constraints. The emergence of the filaments can be seen in the progression from small to large $\beta$s. However, we make the conservative interpretation that the filaments do not become relatively long, narrow, and radial enough until $\beta \gtrsim 400$. Furthermore, the ejecta material in the filaments seen in \autoref{fig: ballistic fraction slices} are moving close to $k\approx 1.0$ and the value of $k$ is more evenly distributed for $\beta = 400$ and $\beta = 800$. 

We will choose $\beta = 800$ to dynamically match with Pa 30. We use $t = \num{.1} \, \tsed$ that corresponds to $\rRS \approx 0.12$, $\rFS \approx 0.20$. For physical units, we will use results from \citealt{cunningham_expansion_2025} as $\rRS \approx \qty{0.6}{pc}$, $\rFS \approx \qty{1.25}{pc}$. We will use an age corresponding to 1181 AD as $\qty{844}{yr}$. 

This choice is further backed by the illustration of the velocity sampling as a function of radius in \autoref{fig:scatter plot}. At the $t=0.1 \, \tsed$, we demonstrate a scatter plot of a sampling of points in our domain colored by the ejecta fraction. We note that the velocity range is $\approx 700-1300 \unit{\km\per s}$, which matches quite closely to the velocity range measurements presented in \citealt{cunningham_expansion_2025}. Furthermore, we see that the shocked region is conservatively captured by our tracking conditions discussed for the reverse and forward shock in \autoref{methods subsec: tracking rRS and rFS}.

Now that we have a dynamical time identified for Pa 30, we use our results to discern the CSM number density $n_{\csm}$ and initial explosion energy $\Eej$ in terms of the ejecta mass $M_{\ej}$, age $t$, and its forward shock radius $\rFS$. 

\begin{equation} \label{eq: nCSM}
    \rho_{\csm} \approx \qty{1.47 e-25}{\gram\per\cm\cubed} \times \left(\frac{\Mej}{0.1 \, \Msun}\right)\left(\frac{\rFS}{\qty{1.25}{pc}}\right)^{-3}
\end{equation}

\begin{equation}
    n_{\csm} \approx \frac{\rho_{\csm}}{m_p} \approx \qty{0.088}{\per\cubic\cm}
\end{equation}

\begin{equation}\label{eq: E_Pa30}
    \Eej \approx \qty{3.5e47}{erg} \times \left(\frac{r_{\FS}}{\qty{1.25}{pc}}\right)^2 \left(\frac{t}{\qty{844}{yr}}\right)^{-2} \left(\frac{M_{\ej}}{0.1 \, \Msun}\right)
\end{equation}

We see that in order to create the filamentary structure seen in Pa 30, the energy required in the initial explosion is much less than a typical supernova ($\gtrsim 1000$ times less than a typical SN Ia explosion). This would support the theory that SN 1181 was a Type Iax that followed a disruptive WD-WD merger, since these circumstances would create a sub-luminous transient (e.g. \citealt{foley_type_2013}). 

In \autoref{results subsec: Lcool}, we discussed that to create the range of morphological structures, the cooling mechanism doesn't remove more than $1 \%$ of the ejecta energy for any of the $\beta$ runs at the relative age of Pa 30. Using our result in \autoref{eq: E_Pa30}, we can estimate the cooling luminosity in terms of an efficiency $\epsilon (0 <\epsilon < 1)$, which we scale by $0.01$ since this was the upper limit we found for our model suite in \autoref{fig:cooling luminosity}.

\begin{equation} \label{eq: Lcool}
    L_{\text{cool}} \approx \qty{e35}{erg\per\second} \times \left(\frac{\epsilon}{0.01}\right) \left(\frac{\Eej}{\qty{3.5e47}{erg}}\right) \left(\frac{t}{\qty{844}{yr}}\right)^{-1}
\end{equation}

If the observed total luminosity of Pa 30 is comparable to this number in any band, this is sufficient to explain its morphology via radiative cooling.

\subsection{Cosmic Ray Emission?} \label{discussion subsec: CRs and Tycho}

There is evidence that suggests the Tycho SNR emits a portion of its thermal energy into cosmic rays (e.g. \citealt{warren_cosmicray_2005}). In Tycho, imaging has shown that parts of the ejecta have caught up with the forward shock and pierced through the interface. Efficient cooling from the forward shock via cosmic ray emission could explain this property of Tycho. Perhaps cosmic ray emission is active in Pa 30. However, the environments and structure of the ejecta are very different. Tycho's ejecta takes on puffy, cauliflower shapes while Pa 30's ejecta is morphed into long and narrow spikes. We also note that Pa 30 can be simulated under rapid radiative cooling that results in a weaker and corrugated shock. Cosmic ray emission is dependent on a strong shock and modeling emission in Pa 30 would require different physics than for Tycho.

Our study removes a constant thermal energy throughout the whole system and with time. In reality, cosmic ray emission would have time and spatial dependencies in correlation to the strength of the shock. Future work could entail specifically modeling strong cooling via cosmic ray emission.

\subsection{Precise Measurement of Expansion Velocities}
Although efficient cooling results in nearly ballistic expansion speeds, the ejecta is still decelerated in the process, which predicts a ballistic fraction $k \lessapprox 1$. If the strong winds created from the spinning compact object are sufficient enough to interact with the material far into the shocked region, we would expect the ejecta to be accelerated from free expansion velocities, thus having a $k > 1$. Based on \citealt{cunningham_expansion_2025}, $k\approx1$ which falls on the threshold between the predictions for cooling-dominated or wind-driven models. We propose that further constraints on the ballistic fraction of the ejecta by more precise measurements could further distinguish between the effects of cooling or wind in the creation of the filamentary structure in Pa 30. Perhaps, using imaging techniques at different epochs could provide another way to measure the expansion velocities (e.g. \citealt{martin_non-uniform_2025}). 

\begin{figure}
    \centering
    \includegraphics[width=\linewidth]{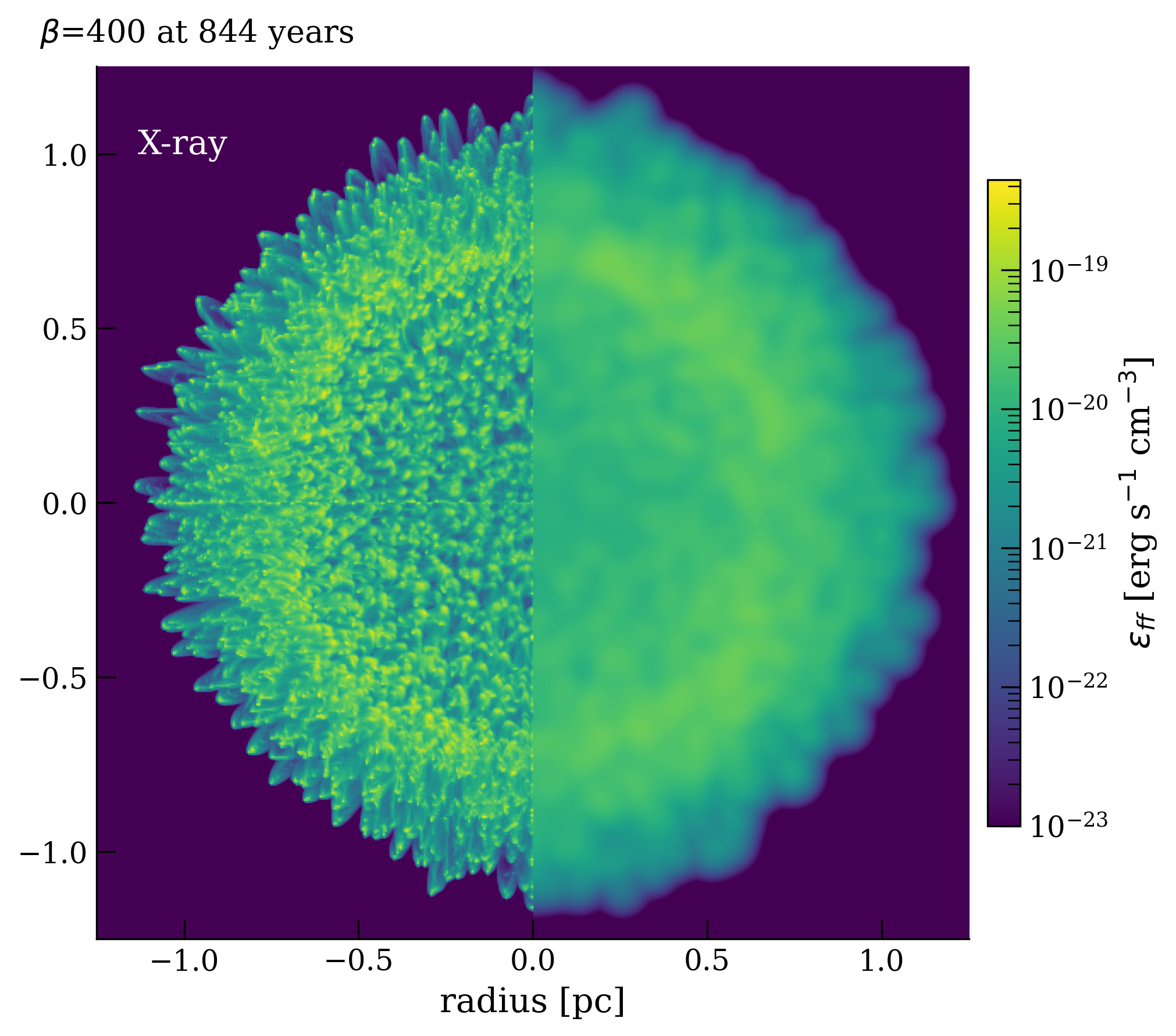}
    \caption{Synthetic X-ray emissivity map using \autoref{eq: xray emissiv} for $\beta=400$ and $t/\tsed=0.1$. The left hand side is the integrated emissivity extrapolated to half a circle. The right side is a gaussian smoothed version of the left side with variance $\sigma=6$ pixels, which corresponds to a resolution of $\approx \qty{2}{\arcsecond}$.}
    \label{fig:xray map}
\end{figure}


\subsection{Synthetic X-ray map}
The Pa 30 nebula is observed in X-rays, providing insight to the explosion and emission mechanisms (\citealt{gvaramadze_massive_2019}, \citealt{oskinova_x-rays_2020}). In order to further test the prediction of efficient radiative cooling in the ejecta, we calculate approximate X-ray emissivity maps using one of the models in our suite.

For simplicity, we assume the X-ray emission is due to thermal Bremsstrahlung and calculate an approximate emissivity (e.g. see \citealt{rybicki_radiative_1986}). We estimate the temperature assuming an ideal gas and a composition fully composed of Sulfur. Given a temperature $T$, electron number density $n_e$, ion number density $n_i$, atomic number for composition $Z$, and a velocity averaged Gaunt factor that we approximate as $\bar{g}_B \approx 1.2$, we can calculate the thermal X-ray emissivity in each grid zone of our hydrodynamical models.

\begin{equation}\label{eq: xray emissiv}
    \varepsilon_{ff} \approx (\qty{1.4e-27}{erg\per\cubic\cm}) T^{1/2} n_e n_i Z^2 \bar{g}_B(T)
\end{equation}

With the thermal X-ray emissivity approximations, we integrate along the line of sight to create the synthetic maps in \autoref{fig:xray map} for $\beta=400$ and $t/\tsed = 0.1$. Furthermore, we include a smoothed version our X-ray map data using a gaussian convolution, with a variance of $\sigma = 2$ arcseconds in order to approximate the blurring in X-ray imaging of the nebula due to resolution constraints of the instrument.

\begin{figure*}
    \centering
    \includegraphics[width=\linewidth]{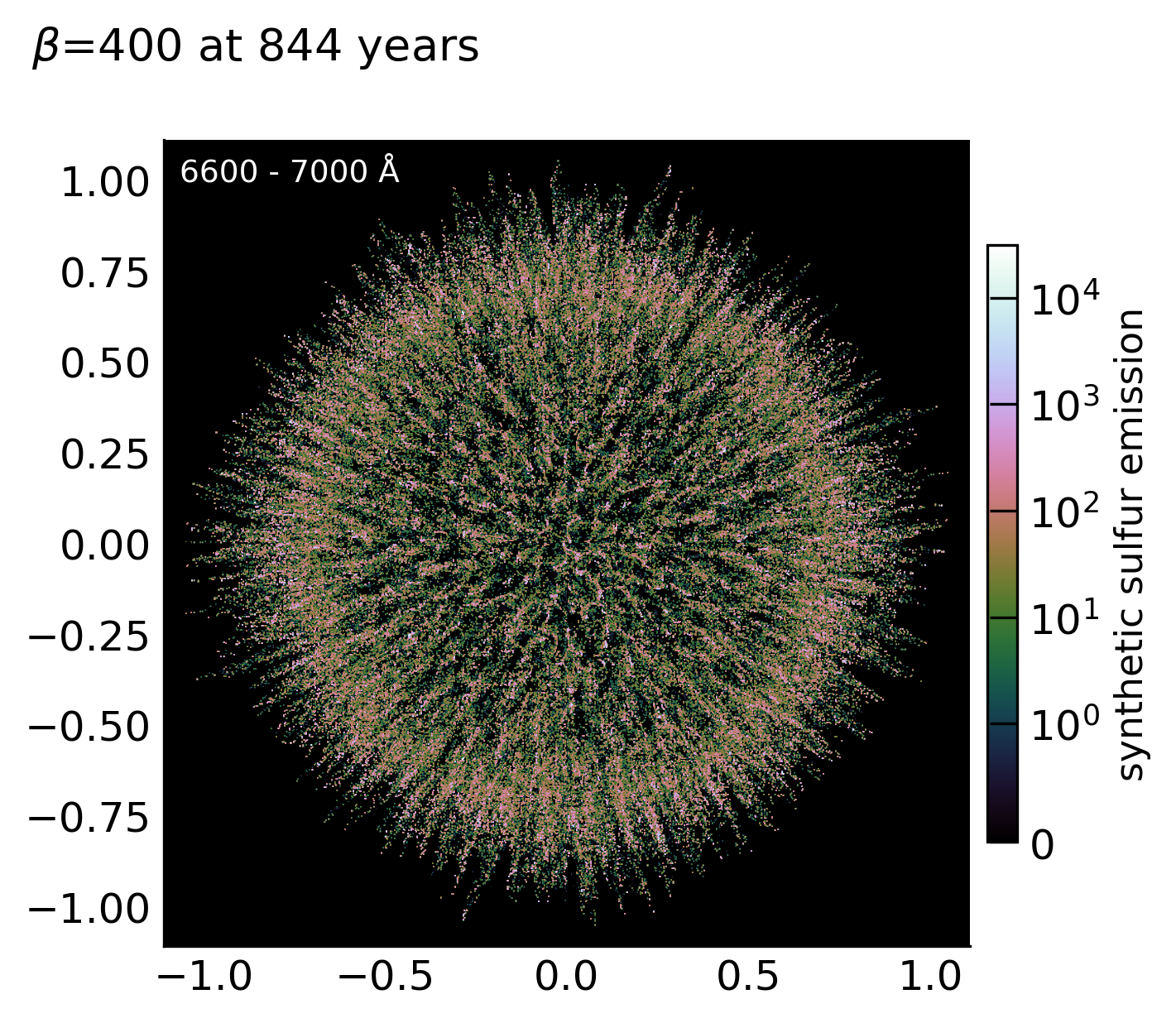}
    \caption{Synthetic \ion{S}{2} emissivity maps for $\beta=400$ and $t/\tsed=0.1$ using \autoref{eq: line cooling emissiv} and integrated along the line of sight in approximate emitting regions for \ion{S}{2}.}
    \label{fig:SII emission map}
\end{figure*}

\subsection{Predicted [SII] emission maps}

We also calculate synthetic sulfur \ion{S}{2} emissivity maps for $\beta=400$ at $t/\tsed = 0.1$. We assume an ideal gas law to calculate the temperature and integrate a line cooling emissivity along the line of sight only in the zones falling within a relevant wavelength range for \ion{S}{2} emission. In other words, we compute the total emissivity by a summation only where $kT/h$ is in the sulfur band, given Boltzmann constant $k$, temperature $T$, and Planck constant $h$. For density $\rho$ and passive scalar (ejecta fraction) $X$, we estimate a line cooling emissivity for the $z$-axis as,

\begin{equation}\label{eq: line cooling emissiv}
\varepsilon_{\text{line cooling}} \approx \int_{\text{is emitting}} \rho^2 X \, dz
\end{equation}

The ``is emitting" criterion in our integration refers to the temperature range for \ion{S}{2} emission that we discuss in the previous paragraph. The estimated \ion{S}{2} emissivity is pictured in \autoref{fig:SII emission map} where we choose a relevant wavelength range between \qty{6600}{\angstrom} and \qty{7000}{\angstrom}. We rotate and flip our octant data cube to extrapolate our results to an image of the entire nebula. There is remarkable similarities in our synthetic map to imaging taken in \citealt{cunningham_expansion_2025} and \citealt{fesen_discovery_2023}. The temperature selection and projection effects create the desired radial and filamentary structure in Pa 30. We note that the stacking of filaments on top of each other creates the appearance of clumps or clusters of emitting zones along the filaments.

\subsection{Observed Radio Emissions?}

Imaging the radio and infrared emission from the nebula could further inform our predictions of strong cooling in Pa 30.  In \citealt{shao_absence_2025}, the radio surface brightness measured in Pa 30 was at least $1000$ times less than a typical supernova remnant with upper flux density limits of $\qty{0.84}{mJy}$ in 1.5 GHz and $\qty{0.29}{mJy}$ in 6 GHz.

One possible way to explain the faint radio signal is the predicted low shock pressure induced by rapid cooling, portrayed in \autoref{fig:pressure slices}. For example, a remnant with $~800$ times faster cooling mechanism would be orders of magnitude fainter than the no cooling case. If the the shock was able to be resolved with deep observations, the resulting morphology and measurements could provide insight on whether or not cooling is prevalent in Pa30. According to the predictions of this work, the forward and reverse shocks would be hard to identify since they would be highly corrugated and order of magnitude cooler (see \autoref{fig:pressure slices}). The deviation from a spherically smooth and obvious signatures of the forward shock that we typically observe in remnants could explain why Pa30 is not being detected in the radio.

\subsection{Computating Line Emission Cooling Timescales}
Realistically, the composition of the evolving supernova remnant will be important for modeling the cooling scenario. The stratification of different atomic elements within the ejecta will vary the cooling rates spatially and different sections will be cooled off faster than others. This work aims to generically cool of supernova remnants in order to explore the parameter space in an unbiased way. However, important and relevant future work would entail modeling the cooling in more detail by adding dependence of composition. For Pa 30, a system that appears to be rapidly cooling off, the composition and presence of heavier metals could play a crucial role in understanding its physics. In fact, significant infrared emission in Pa 30 suggests the presence of cold dust (\citealt{lykou_new_2023}). Observations and theoretical modeling predict Type Iax supernovae should enrich their surrounding environments with dust (e.g. \citealt{nozawa_formation_2011}, \citealt{fox_excess_2016}, \citealt{kumar_type_2025}). Sufficient amounts of dust could drastically affect the radiative cooling losses and should be modeled with care. Effects of dust and chemical composition are not within the bounds of this study but provide important context for white dwarf explosions like Type Iax supernovae.

Nonetheless, we compute a physical cooling timescale for one of our models to Pa 30 in order to provide observational context within the regime of our parameter study. We assume that the power emitted via line cooling is related by a cooling rate coefficient $\Lambda$, ion number density $n_i$ and electron number density $n_e$ (see e.g. \citealt{raymond_shock_2018}). The cooling coefficient $\Lambda$ is typically computed with atomic data calculations and will depend on the electron temperature and composition of the material.

\begin{equation}
    P_{\text{line cooling}} = \Lambda(T_e) \, n_i \, n_e \,\, \unit{erg \per \cubic \cm \per \s}
\end{equation}

We use this formalism to compute an approximate cooling timescale via line cooling for our model suite. The cooling timescale $\tcool$ is found by calculating the total thermal energy in the shocked region and the total luminosity due to line cooling given a choice for $\Lambda$, atomic number of the ejecta $Z$, and atomic mass number of the ejecta $A$.

\begin{gather}
    \tcool \approx \frac{\int_{\text{shocked}} \varepsilon_{th} \, dV}{\int_{\text{shocked}} \Lambda n_e n_i \, dV} \\ \approx  \frac{\int_{\text{shocked}} \varepsilon_{th} \, dV}{\Lambda \frac{Z}{A^2}\frac{1}{m_p^2} \int_{\text{shocked}}\rho^2 \, dV}
\end{gather}

We report the cooling time for our $\beta = 800$ model below:

\begin{equation}
    \tcool \approx \num{200}\unit{yr} \times \left(\frac{\Lambda}{\qty{e-19}{erg\per\cubic\cm\per\s}}\right)^{-1} \left(\frac{Z}{16}\right)^{-1} \left(\frac{A}{32}\right)^{2} 
\end{equation}

This cooling timescale of 200 years is slower than the $\sim 20$ year timescale that was needed in our model suite to produce filaments. Thus, using $\Lambda \sim 10^{-19}$ (Typical for metal-rich ejecta) is insufficient by itself to produce the filamentary structures around $\beta \gtrsim 400$. These results suggest that other mechanisms besides metal line cooling are needed to explain Pa 30. For example, if the remnant is highly contaminated with dust, the combined contributions from both dust and metals might provide sufficient cooling for the ejecta.  Future JWST observations should help to constrain the dust and heavy metal composition of Pa 30.

\subsection{Are filaments uniformly distributed or randomly distributed?}

\begin{figure}
    \centering
    \includegraphics[width=\linewidth]{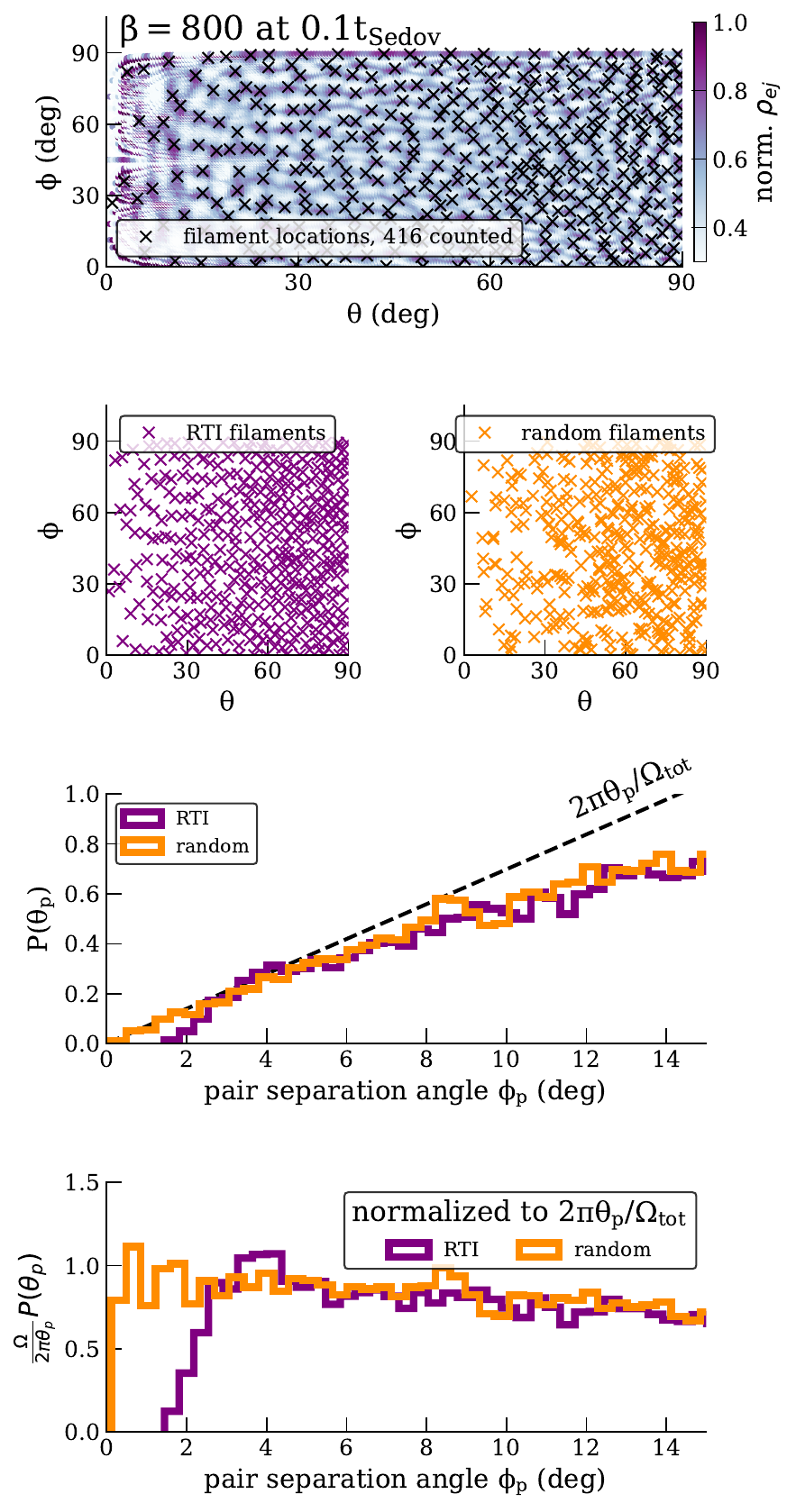}
    \caption{\textit{First row}: normalized ejecta density binned by angular coordinates $\phi$ and $\theta$ at $t=0.1 \, \tsed \approx \qty{844}{yr}$ for $\beta = 800$. The darker purple color denotes a higher density region, thus mapping the locations of the filaments in angular space. The black crosses denote the location of the filaments and are plotted again as purple crosses in the left panel of the second row. \textit{Second row}: locations of filaments generated by RTI (this work and shown in the top panel as well) versus as randomly generated locations shown as orange crosses in the right panel. \textit{Third row}: histogram for the angular separation between a pair of filaments. An analytical prediction for small separation angle (\autoref{eqn:random_fil}) is drawn as a black dashed line. \textit{Fourth row}: The same histogram in the third row but normalized to the analytical formula. We demonstrate the differences in angular separations between a uniform and random distribution of filaments. We point out that if the filaments are generated at random locations (e.g. due to randomly placed clumps), there will be an overabundance of separations less than $2-4$ degrees that is not predicted in our rapid cooling model whose filaments are driven by RTI.}
     \label{fig:filament seps}
\end{figure}

The filamentary structure naturally driven by RTI results in a distribution that is much more uniform than would be expected of a mechanism relating to random clumps or bullets of material.  This is because RTI sets a characteristic angular separation to the filaments.  This can be measured directly if one can determine all of the angular positions of some subset of the filaments.  

A visualization and corresponding analysis of the angular separation between filaments due to RTI vs. randomly generated locations are presented in \autoref{fig:filament seps}. The second row shows the immediate visual difference between the two scenarios. For randomly generated filaments, which could be created by randomly placed clumps of material, some subgroups of the filaments are clustered together. On the other hand, the filaments from RTI in this work are more uniformly spaced. 


We demonstrate this rigorously by computing the correlation function, $dN_{\rm pairs}/d\theta_p$, which can be determined very easily if one knows the $(\theta,\phi)$ positions of a sufficiently large number of filaments.  The correlation function is computed simply by identifying each pair of filaments and computing the angular separation $\theta_p$ of the pair.  If filament 1 is at position $(\theta_1,\phi_1)$ and filament 2 is at $(\theta_2,\phi_2)$, their angular separation is simply given by the dot product of their corresponding unit vectors:

\begin{align}
    {\rm cos} ~\theta_p & = {\rm cos}~\theta_1 {\rm cos}~\theta_2 \\ & + {\rm sin}~\theta_1 {\rm cos}~\phi_1 {\rm sin}~\theta_2 {\rm cos}~\phi_2 \\ & + {\rm sin}~\theta_1 {\rm sin}~\phi_1 {\rm sin}~\theta_2 {\rm sin}~\phi_2
\end{align}

Then, all pairs are binned into a histogram, so that the number of pairs $dN_{\rm pairs}$ per bin width $d\theta_p$ is $\frac{d N_{\rm pairs}}{d\theta_p}$.  This can then be normalized by the total number of pairs, so that the entire curve integrates to unity:

\begin{equation}
    P(\theta_p) \equiv \frac{1}{N_{\rm pairs}} \frac{d N_{\rm pairs}}{d\theta_p}.
\end{equation}

This correlation function is most useful for small values of $\theta_p$ (at large distances, filaments are essentially uncorrelated).  For a purely random distribution, the number of pairs with a distance $\theta_p$ from a filament should be proportional to the number of filaments per unit solid angle, $\frac{dN_f}{d\Omega}$:

\begin{equation}
    N_{\rm pairs}[\theta < \theta_p] = N_f \frac{dN_f}{d\Omega} \pi \theta_p^2
\end{equation}

(again, assuming small $\theta_p$).  Therefore,

\begin{equation}
    \frac{d N_{\rm pairs}}{d\theta_p} = N_f \frac{dN_f}{d\Omega} 2 \pi \theta_p = N_{\rm pairs} \frac{2 \pi \theta_p}{\Omega_{\rm tot}}.
\end{equation}

Here, $\Omega_{\rm tot}$ is the total solid angle subtended by our sample of filaments.  The entire remnant has $\Omega_{\rm tot} = 4 \pi$, but in our case we are only evaluating an octant, for which $\Omega_{\rm tot} = \pi/2$.

Therefore, for a purely random distribution of filaments, one expects

\begin{equation}
    P(\theta_p) \rightarrow 2 \pi \theta_p/\Omega_{\rm tot}, ~{\rm for}~ \theta_p \ll 1.
    \label{eqn:random_fil}
\end{equation}

However, for a roughly uniform distribution of filaments with characteristic separation $\Delta \theta$, one expects a deficit in the distribution for $\theta_p < \Delta \theta$ and a peak at $\theta_p = \Delta \theta$.

We find such a deficit and peak in the correlation function for our distribution of filaments, with a filament separation of $\Delta \theta \approx 4^\circ$ (bottom two panels of \autoref{fig:filament seps}).  We identified 416 filaments in our octant, corresponding to

\begin{equation}
    \frac{dN_f}{d\Omega} \approx 270
\end{equation}

Meaning that the entire remnant would be spanned by about 3000 filaments.  We additionally show that choosing 416 randomly distributed filaments would give a different prediction for this correlation function, with no deficit or peak, following Equation (\ref{eqn:random_fil}) for small $\theta_p$.  If we re-normalize by Eq (\ref{eqn:random_fil}), the difference between our filament distribution and a random one becomes even more prominent:

\begin{equation}
    \frac{\Omega_{\rm tot}}{2 \pi \theta_p} P(\theta_p) = \frac{\Omega_{\rm tot}}{ 2 \pi \theta_p N_{\rm pairs}} \frac{dN_{\rm pairs}}{d\theta_p},
\end{equation}
which tends to unity as $\theta_p \rightarrow 0$ for a random distribution, but exhibits our deficit and peak for a uniform distribution such as that generated by our RTI filaments (bottom panel of \autoref{fig:filament seps}).

\section{Conclusions} \label{section: Conclusion}
In this paper, we used 3D hydrodynamical models of supernova remnants to systematically explore how cooling rates can create structural changes during their evolution. We characterize our SNR model suite by a singular cooling parameter $\beta$, that controls the fraction of the thermal energy that is removed from the system at each time step. Our cooling implementation is not biased towards certain cooling mechanisms and only relies on the thermal energy present in the system. Essentially, $\beta$ is a generalized prescription to investigate SNR cooling evolution in a systematic way, which we use to interpret the peculiar filamentary structure of the Pa 30 nebula. 

Future work would include exploring specific cooling mechanisms in depth. Our methods did not account for time-dependence or whether the density, temperature, and velocity of the shock radii impacts the behavior of the cooling. We also acknowledge that the Pa 30 nebula hosts a rapidly spinning object that produces extreme wind speeds. We replicate Pa 30's filamentary structure without modeling the fast winds, considering that the reverse shock radius in Pa 30 is $\sim \qty{0.6}{pc}$ whereas the wind termination shock is suggested to be confined to $< \qty{0.02}{pc}$ (\citealt{ko_dynamical_2024}), but it is possible that the extent of the wind is larger than this estimate. Perhaps ionization effects of the wind and magnetic fields play a crucial role. 

Ultimately, we demonstrated that one can produce a range of SNR morphologies by only increasing or decreasing a generic cooling timescale in the material. Our methods allow us to characterize SNRs with a singular value $\beta$ perhaps signifying all remnants lie on a cooling spectrum. We summarize the key results from this paper below:
\begin{itemize}
    \item  Increasing $\beta$ causes the RTI fingers in the ejecta to grow into long radial spikes. We explain this effect as efficient cooling limits the effects of the Kelvin-Helmholtz instability, thus preventing the heads of the RTI fingers from turning over. 
    \item A progression of ejecta morphologies is produced when analyzing the structure with no cooling to our most rapid cooling case. The morphologies evolve from the common puffy, cloud-like structures seen in Tycho to the elongated, narrow, and radial spikes observed in Pa 30.
    \item Significant cooling slows down the expansion speeds of the remnant, allowing the RTI fingers in the shocked region to catch up with the forward shock. This results in a heavily corrugated forward shock. 
    \item For $\beta \gtrsim 400$, the RTI fingers are sculpted into many long and narrow filaments. The ejecta in these filaments have not swept up enough mass to effectively decelerate and is still moving close to free expansion speeds. Specifically, we measured that the ballistic fraction $k = v/(r/t) \gtrsim 0.9$ for $\beta = 400$ and for $\beta = 800$. 
    \item We analyze the time-dependence of how much energy was being cooled off in our systems and equated this to a cooling luminosity. Across all $\beta$ values in our model set, the cooling luminosity does not exceed $2\%$ of the total ejecta energy at any time. 
    \item By using the ratio of the reverse to forward shock radius and a requirement that the ejecta must be traveling at least $\approx 90\%$ their free expansion speed, we dynamically match $\beta = 800$ to Pa 30 to estimate an ejecta energy of $\approx \qty{e48}{erg}$ with an associated cooling luminosity $\approx \qty{e36}{erg\per\s}$. 
    \item We make observational predictions for the Pa 30 nebula that could support or argue against the predictions of strong cooling. Our results predict pressure magnitudes that are magnitudes smaller under strong cooling, suggesting that identification of the forward shock in the radio would be very difficult. We also compute synthetic X-ray and Sulfur maps that show similarities to imaging of the nebula.
    
    \item We count about 400 filaments for an octant (extrapolated to about 3000 filaments for the entire remnant) and perform statistical analysis on the angular separations between filaments. This demonstrates that randomly distributed filaments will cluster more at smaller separations as opposed to a more uniform distribution generated from RTI with efficient cooling.
\end{itemize}
Based on these findings, we look forward to and recommend future additional observations of Pa 30 to confirm or deny cooling as the source of its morphology. 
\begin{acknowledgments}
\begin{center}
\textbf{Acknowledgments} 
\end{center}
We would like to thank Tim Cunningham and Maxim Lyutikov for helpful comments and suggestions. In addition, we thank Dillon Hasenour for comments on early drafts of the manuscript. We also thank Andreas Ritter for helpful comments on relevant literature used. We thank Dan Milisavljevic for comments and discussion on radiative cooling in Pa 30. Lastly, we thank Takatoshi Ko for insight and aide with understanding X-ray emission modeling and observations. Calculations were carried out using the Petunia computing cluster hosted by the Department of Physics and Astronomy at Purdue University, hence we thank Chris J. Orr for maintaining Petunia.

\textit{Software}: \texttt{Sprout} \citep{mandal_sprout_2023}, VisIt \citep{childs_high_2012}
\end{acknowledgments}

\appendix

\section{Density Perturbation Seeding in the Circumstellar Environment}\label{append: pert seed}

In this section, we investigate the effect of the mixing layer between the reverse and forward shock (and therefore the development of the Rayleigh-Taylor filaments) as a function of the perturbation seed that we use in initializing the surrounding environment density. 

\begin{figure}
    \centering
    \includegraphics[width=\linewidth]{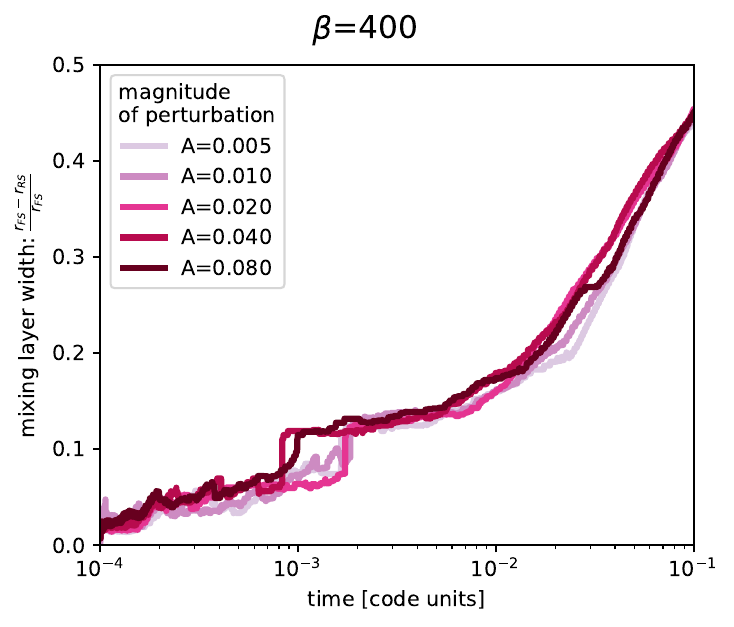}
    \caption{The mixing layer width with time for different perturbation strengths. The perturbation magnitude is determined by the value of $A$ which is described in \autoref{methods subsec: initial conditions}. The mixing layer width does not substantially differ across different perturbation seeds, demonstrating that our choice for $A$ does not affect our interpretations of the RTI filaments and structure in this work. }
    \label{fig:csm perturbation}
\end{figure}

In \autoref{fig:csm perturbation}, we plot $\frac{\rFS - \rRS}{\rFS}$ as a function of time for different perturbation magnitude values of $A$. This provides a proxy for the mixing layer width and therefore the behavior and emergence of the RTI filaments that form at in the mixing layer of the evolving supernova remnant. 

For completeness, we again present the density perturbation we implemented that is also described in \autoref{eq: density perturb}. 

\begin{equation} 
    \rho_{\csm}  = 1 + A \sin{k\phi} \sin{k\theta}
\end{equation} 

\autoref{fig:csm perturbation} demonstrates that the mixing layer width temporal evolution is independent of the initial perturbation seed that we add to the environment density. Therefore, this section serves to suggest that the emergence of the filaments and the subsequent kinematic and structure study that we perform in this paper is not due to our choice of the initial density conditions. 

\section{Wind-Blown Circumstellar Environment}

A uniform circumstellar density environment is not representative of typical and realistic environments for supernovae. In reality, the surroundings would be enriched with material that changes the density such as precursor mass expulsions, dust, or winds ejected by the progenitor star. 

In this section, we explore the filament creation in response to a different CSM density profile. The main goal is so test if the filamentary structure that is created at our radiative cooling threshold of $\beta \sim 400$ is substantially affected by the surrounding medium density.  We model the wind-like CSM as a simple power law:

\begin{equation}
    \rho_{\csm} \sim \frac{1}{r^2}
\end{equation}

\begin{figure} 
\centering
    \includegraphics[width=\linewidth]{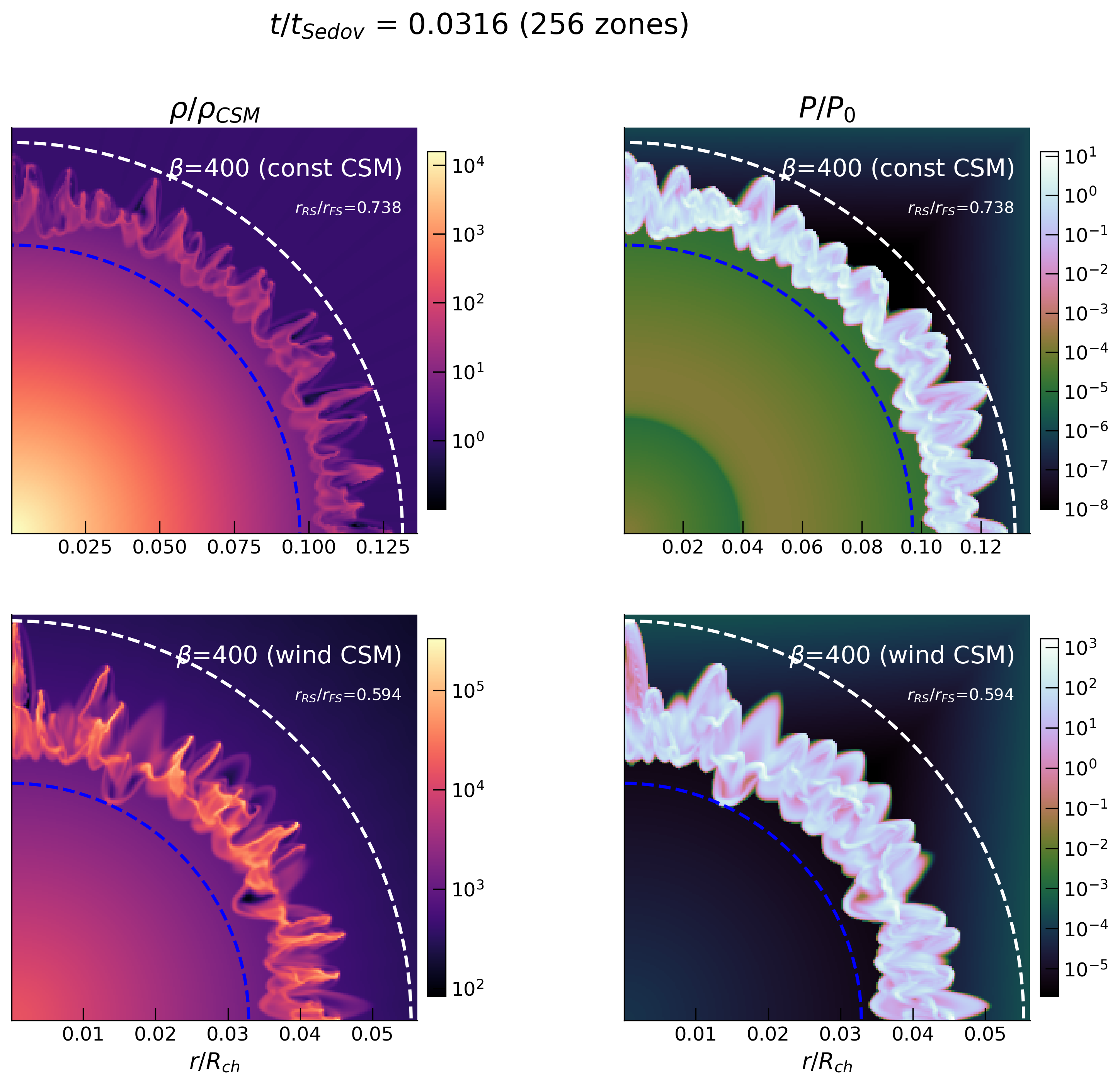}
    \caption{Density (left panels) and pressure slices (right panels) at $t/\tsed \approx .03$ for two different initial CSM density profiles with $\beta = 400$. Quantities are presented in unit-less quantities for simple comparison and due to the fact that changing the initial CSM density would require different physical scalings than the ones used in this work (see \autoref{table: runtime params}). The wind-like CSM case evolves faster, which is expected as the forward shock sweeps up more mass at earlier times.}
    \label{fig: wind CSM compare}
\end{figure}

In \autoref{fig: wind CSM compare}, we plot density and pressure slices for the uniform CSM density and a wind-like CSM density with a cooling prescription set to $\beta = 400$. We choose to display the density and pressure at $t/\tsed \approx .03$ because the reverse to forward shock ratio in the wind-like CSM case is $\rRS/\rFS \approx 0.6$. Recall that for the uniform case (and this work), when the shock ratio condition reaches this value, $t/\tsed = 0.1$. This demonstrates that the wind-like CSM density causes the SNRs to evolve quicker. This can be explained by the fact that a steep density drop off means more mass is concentrated at smaller radii, meaning that the forward shock sweeps up more mass at earlier times compared to a uniform CSM case. We note that the total CSM mass is the same for both the constant density case and the wind-blown case; therefore, this result cannot be attributed to there simply being more material. Nonetheless, the RTI filaments are still structured similarily to the results we have shown for the constant CSM scenario. This suggests that more realistic density profiles for the surrounding medium will affect the time evolution of the SNRs, but the overall structure and behavior of RTI will not be affected substantial in regards to the motivations of this study. Future studies would benefit from more detailed CSM density models, but we argue that the behavior of RTI in varying degrees of strongly cooled ejecta will be comparable. 

\bibliography{references}{}
\bibliographystyle{aasjournal}

\end{document}